\documentclass[prd,aps,preprint,showpacs,nofootinbib,superscriptaddress,tightenlines]{revtex4-1}
\usepackage{epsfig}
\usepackage{amssymb}
\usepackage{amsmath}
\usepackage{graphicx}
\usepackage{psfrag}

\begin{document}

\title{Universality of Unintegrated Gluon Distributions at small x}
\author{Fabio Dominguez}
\affiliation{Department of Physics, Columbia University, New York,
NY, 10027, USA}
\author{Cyrille Marquet}
\affiliation{Physics department, Theory Unit, CERN, CH-1211 Geneva,
Switzerland}
\author{Bo-Wen Xiao}
\affiliation{Department of Physics, Pennsylvania State University, University Park, PA 16802, USA}
\affiliation{Nuclear Science Division, Lawrence Berkeley National
Laboratory, Berkeley, CA 94720, USA}
\author{Feng Yuan}
\affiliation{Nuclear Science Division, Lawrence Berkeley National
Laboratory, Berkeley, CA 94720, USA}
\affiliation{RIKEN BNL Research
Center, Building 510A, Brookhaven National Laboratory, Upton, NY
11973, USA}

\begin{abstract}
We systematically study dijet production in various processes in
the small-$x$ limit and establish an effective $k_t$-factorization
for hard processes in a system with dilute probes scattering on a
dense target. We find that the well-known Weizs\"{a}cker-Williams gluon distribution can be directly probed in the quark-antiquark jet correlation in deep inelastic 
scattering and the dipole gluon distribution can be directly measured in the direct photon-jet correlation in $pA$ collisions.  In the large-$N_c$ limit, the unintegrated gluon
distributions involved in other different dijet channels in pA collisions are shown to be
related to two widely proposed ones: the Weizs\"{a}cker-Williams
gluon distribution and the dipole gluon distribution.
\end{abstract}

\maketitle

%\preprint{LBNL-xxx}\preprint{RBRC-xxx}

%\email{bxiao@lbl.gov}

%\email{fyuan@lbl.gov}

\section{Introduction}

Factorization is part of the foundations of high-energy hadronic
physics, as it provides the key ingredient for the
phenomenological studies of high-energy experiments. Factorization
theorems make the separation between short-distance perturbative
physics and long-distance nonperturbative effects possible. Thus,
cross sections measured in high-energy experiments can be
factorized into products of hard parts (short-distance physics)
and parton distributions (nonperturbative physics). In addition,
an essential part of factorization is the universality of the
parton distributions, among different processes.

While collinear factorization has been the most widely used
framework in phenomenological studies, and remains a sufficiently
good approximation of QCD for the most inclusive processes in
hadronic collisions, the investigation of less inclusive
observables showed the need for a transverse-momentum dependent
(TMD) factorization. During the last decade, a large amount of
work has been devoted to establish such a framework in QCD.
However, recent progress
\cite{BoeVog03,Bomhof:2006dp,qvy-short,Collins:2007nk,Vogelsang:2007jk,Rogers:2010dm,
Xiao:2010sp} has shown that TMD factorization is violated for
dijet production in hadron-hadron (e.g., pp) collisions, due to a
loss of universality.

In this paper, we propose a solution to this problem in the
small-$x$ limit: we succeeded in establishing an effective TMD
factorization for hard processes in collisions of dilute probes
off dense hadrons (or large nuclei)\footnote{Note that this effective TMD factorization does not hold for high energy pp and AA collisions due to final state soft gluon exchanges from both the projectile and the target to the hard part. See Ref.~\cite{Rogers:2010dm} for detailed discussion. For the case of dilute projectiles scattering on a dense target, we can always neglect the soft gluon exchanges from the dilute projectile to the hard part while we resum all the soft gluon exchanges attached the dense target to the hard part since the gluon field is much stronger in the dense target.}. We confirm that TMD parton
distributions are not universal, but we show that at small-$x$
they can be constructed from several universal individual building
blocks. This is achieved by working with an appropriate
approximation of QCD in the small-$x$ limit of QCD, where large
parton densities and non-linear {\it saturation} effects are
crucial.

The saturation phenomena in high-energy collisions has attracted
great attention in the last two decades. At very high energies
corresponding to the low-$x$ regime, parton distributions reach
very high densities and non-linear effects become important in
describing the dynamics of the hadronic system
\cite{Gribov:1984tu,Mueller:1985wy,McLerran:1993ni,Iancu:2003xm}.
The transition to the saturation regime is characterized by the
saturation scale, which is interpreted as the typical transverse
momentum of the small-$x$ partons, and is also related to the
transverse color charge density in the infinite momentum frame of
the dense target. It has been argued \cite{McLerran:1993ni} that
the high density of gluons inside a hadron or nucleus allows for a
semiclassical treatment of the color field, leading to the Color
Glass Condensate (CGC) effective description of the small-$x$ part
of the hadronic/nuclear wave function which has been widely used
to systematically study saturation physics \cite{Iancu:2003xm}.

Experimental data is still not conclusive in this matter, but
strong evidence of these effects have been found in the deep
inelastic scattering (DIS) experiments at HERA and deuteron-gold
collisions at RHIC \cite{Iancu:2003xm}. It is expected that
saturation physics will play an important role in explaining the
results of the ongoing measurements of single-inclusive production
and two-particle correlations in the forward region at RHIC as
well as future heavy-ion experiments at LHC. In addition, the
planned Electron-Ion Collider \cite{Deshpande:2005wd} will be able
to provide ideal experimental conditions to study the low-$x$
parton distributions and thus test the saturation physics in both
protons and large nuclei.

In saturation physics, two different unintegrated gluon
distributions (UGDs) have been widely used in the literature. The
first gluon distribution, which is known as the
Weizs\"{a}cker-Williams (WW) gluon distribution, is calculated
from the correlator of two classical gluon fields of relativistic
hadrons (non-abelian Weizs\"{a}cker-Williams fields)
~\cite{McLerran:1993ni, Kovchegov:1998bi}. The WW gluon
distribution has a clear physical interpretation as the number
density of gluons inside the hadron in light-cone gauge, but is
not used to compute cross sections. On the other hand, the second
gluon distribution, which is defined as the Fourier transform of
the color dipole cross section, does not have a clear partonic
interpretation, but it is the one appearing in most of the
$k_t$-factorized formulae found in the literature for
single-inclusive particle production in $pA$
collisions~\cite{Iancu:2003xm}.
%FY
% (here $k_t$-factorization refers to a small-$x$ result and involves an all-twist unintegrated gluon %distribution, it should not be confused with the TMD factorization which is a leading-twist result not %restricted to small $x$).

It was a long-standing question what is fundamentally different between these two gluon distributions, and whether there is any observable sensitive to the WW distribution \cite{Kharzeev:2003wz}. The objective of this paper is to answer these questions and show that these two gluon distributions are the fundamental building blocks of all the TMD gluon distributions at small $x$. Eventually, this leads us to an effective TMD-factorization for dijet production, in the collision of a dilute probe with a dense target. We find that, in the small momentum imbalance limit described below, the dijet production process in DIS can provide direct measurements of the WW gluon distribution and the photon-jet correlations measurement in $pA$ collisions can access the dipole gluon distribution directly. In addition, other more complicated dijet production processes in $pA$ collisions will involve both of these gluon distributions through convolution in transverse momentum space, when the large-$N_c$ limit is taken.

%FY
%In this paper, we investigate dijet (or dihadron) production in various processes (DIS dijet, photon-jet %correlations and dijet correlations in $pA$ collisions) as probes of the unintegrated gluon distributions %(UGDs), the building blocks of the CGC formalism.

A short summary of our study has been published in
Ref.~\cite{Dominguez:2010xd}. Here we present the detailed
derivations, and the precise equivalence between the TMD and CGC
approaches, in the overlapping domain of validity, i.e. to leading
power of the hard scale and in the small $x$ limit. In general,
the TMD factorization is valid whatever $x$ is but is a
leading-twist approach, while the CGC is applicable only at small
$x$ but contains all the power corrections. Since the main
objective of this paper is to understand dijet production
processes theoretically, we will put the phenomenological studies
in a future work.

\begin{figure}[tbp]
\begin{center}
\includegraphics[width=7cm]{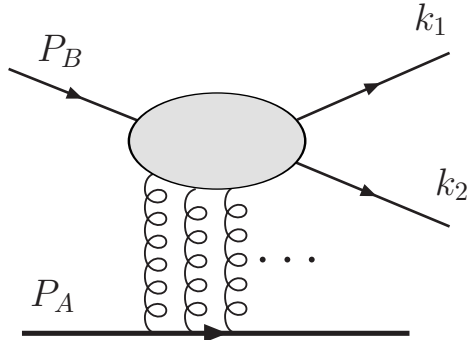}
\end{center}
\caption[*]{Schematic diagrams for two-particle production in a
dilute system scattering on a dense target with multiple scattering.
The imbalance between the two-particle in transverse momentum can be used
to probe the unintegrated gluon distribution of the dense target.}
\label{dijetf}
\end{figure}

We focus on the two particle production (or dijet production at
higher energy) in the case of a dilute system scattering on a
dense target, as illustrated in Fig.~\ref{dijetf},
\begin{equation}
B+A\to H_1(k_1)+H_2(k_2)+X \ ,
\end{equation}
where $A$ represents the dense target (we shall call it a nucleus
in the following), $B$ stands for the dilute projectile (such as a
photon or a high-$x$ parton in a hadron), $H_1$ and $H_2$ are the
final state two particles with momenta $k_1$ and $k_2$,
respectively.  Let us denote as $x_B$ the light-cone momentum
fraction of the parton (or virtual photon) from the incoming
projectile $B$, and as $x_g\ll 1$ the momentum fraction of the
gluon from the incoming target. We are interested in the kinematic
region where the transverse momentum imbalance between the
outgoing particles is much smaller than their individual momenta:
$q_{\perp }=|\vec{k}_{1\perp }+\vec{k}_{2\perp }|\ll k_1\simeq k_2
\simeq P_\perp$ where $\vec{P}_\perp$ is defined as
$(\vec{k}_{1\perp }-\vec{k}_{2\perp })/2$. This is referred to as
the back-to-back correlation limit (the correlation limit) in the
following discussions. An important advantage of taking this limit
is that we can apply the power counting method to obtain the
leading order contribution of $q_{\perp }/P_{\perp }$ where the
differential cross section directly depends on the UGDs of the
nuclei.

For each individual dijet production process, we employ two
independent approaches, namely the TMD approach and the CGC
approach\footnote{Our formulation of the CGC approach leads to
similar intermediate steps as in Ref.~\cite{Nikolaev:2005qs}.
However, we treat the $n$-point functions differently by using
Wilson lines.}. The TMD approach is straightforward and clear in
terms of factorization. On the other hand, the CGC approach is
commonly used in dealing with small-$x$ calculations. It allows to
go beyond the correlation limit, which gives a deeper
access to the QCD dynamics at small $x$, but this is not the
purpose of this paper. In this more general situation, cross
sections involve multi-gluon distribution functions, as expected
due to parton saturation and multiple scatterings, and therefore
there is no $k_t$-factorization. Except for the most inclusive
observables (such as inclusive and semi-inclusive DIS,
single-gluon and valence quark production in pA collisions), $k_t$
factorization is only a property of the linear BFKL regime.
However taking the correlation limit allows to simplify the dijet
production results of the CGC, and to obtain an effective
factorization which coincides with that found in the TMD approach.

%F.Y.
%There have been intensive investigations of these processes in the last few
%years~\cite{BoeVog03,Bomhof:2006dp,qvy-short,Collins:2007nk,
%Vogelsang:2007jk,Rogers:2010dm}, where the associated
%parton distributions are found non-universal. Similarly, the TMD-dependent
%distributions at small-$x$ are also found process-dependent~\cite{Xiao:2010sp}.
%For example, the quark-antiquark jet correlation
%in deep inelastic scattering (DIS) directly probes the first type of the
%UGD, whereas the direct photon-quark jet correlation in $pA$ collisions
%probes the second type of UGD. The dijet (di-hadron) correlations in $pA$
%collisions can probe both gluon distributions, though the connection is more
%complicated. Even though these
%distributions are different for each process, they all can be built from two
%basic building blocks, the Weizs\"acker-Williams distribution and the
%Fourier transform of the dipole cross section.

The Weizs\"acker-Williams gluon distribution can be defined following
the conventional gluon distribution~\cite{Collins:1981uw,{Ji:2005nu}}
\begin{eqnarray}
xG^{(1)}(x,k_{\perp })&=&\int \frac{d\xi ^{-}d^2\xi _{\perp }}{(2\pi
)^{3}P^{+}}e^{ixP^{+}\xi ^{-}-ik_{\perp }\cdot \xi _{\perp }}   \langle P|F^{+i}(\xi ^{-},\xi _{\perp })\mathcal{L}_{\xi }^{\dagger
}\mathcal{L}_{0}F^{+i}(0)|P\rangle \ ,  \label{g1}
\end{eqnarray}%
where $F^{\mu \nu }$ is the gauge field strength tensor $F_a^{\mu \nu
}=\partial ^{\mu }A_a^{\nu }-\partial ^{\nu }A_a^{\mu }-gf_{abc}A_b^\mu
A_c^\nu$ with $f_{abc}$ the antisymmetric structure constants for $SU(3)$,
and
$$\mathcal{L}_{\xi }=\mathcal{P}\exp\{-ig\int_{\xi ^{-}}^{\infty }d\zeta
^{-}A^{+}(\zeta ,\xi _{\perp })\}\mathcal{P}\exp\{-ig\int_{\xi _{\perp
}}^{\infty }d\zeta _{\perp }\cdot A_{\perp }(\zeta ^{-}=\infty ,\zeta
_{\perp })\}$$
is the gauge link in the adjoint representation $A^{\mu}=A_{a}^{\mu }t_{a}$ with $t_{a}=-if_{abc}$.
It contains a transverse gauge link at spatial infinity which is important to make the definition gauge
invariant~\cite{Belitsky:2002sm}. These gauge links have to be made non-light-like to
regulate the light-cone singularities when gluon radiation contributions
are taken into account~\cite{Collins:1981uw}.
%FY
In the above definition, we assume that the hadron is moving along
the $+\hat z$ direction. The light-cone momenta $P^\pm$ are
defined as $P^{\pm}=(P^0\pm P^z)/\sqrt{2}$.
%FY
This gluon distribution can also be defined
in the fundamental representation~\cite{Bomhof:2006dp},
\begin{eqnarray}
xG^{(1)}(x,k_\perp)&=&2\int \frac{d\xi ^{-}d\xi _{\perp }}{(2\pi )^{3}P^{+}}%
e^{ixP^{+}\xi ^{-}-ik_{\perp }\cdot \xi _{\perp }} \langle P|\text{Tr}\left[F^{+i}(\xi ^{-},\xi _{\perp })\mathcal{U}%
^{[+]\dagger }F^{+i}(0)\mathcal{U}^{[+]}\right]|P\rangle \ ,\label{g1fund}
\end{eqnarray}%
where the gauge link $\mathcal{U}_\xi^{[+]}=U^n\left[0,+\infty;0\right]U^n%
\left[+\infty, \xi^{-}; \xi_{\perp}\right]$ with $U^n$ being
reduced to the light-like Wilson line in covariant gauge. It is
straightforward to see that $\mathcal{U}^{[+]}$ represents the
final state interactions according to its future integration path
to $+\infty$.

By choosing the light-cone gauge with certain boundary condition
for the gauge potential ($A_{\perp }(\zeta ^{-}=\infty)=0$ for the
specific case above), we can drop out the gauge link contribution
in Eqs.~(\ref{g1}) and (\ref{g1fund}) and find that this gluon
distribution has the number density interpretation. Then, it can
be calculated from the wave functions or the WW field of the
nucleus target \cite{McLerran:1993ni,{Kovchegov:1998bi}}. Within
the CGC framework, this distribution can be written in terms of
the correlator of four Wilson lines as (see Section \ref{discgc}),
\begin{equation}
xG^{(1)}(x,k_\perp)=-\frac{2}{\alpha_S}\int\frac{d^2v}{(2\pi)^2}\frac{d^2v'}{(2\pi)^2}\;e^{-ik_\perp\cdot(v-v')}\left\langle\text{Tr}\left[\partial_iU(v)\right]U^\dagger(v')\left[\partial_iU(v')\right]U^\dagger(v)\right\rangle_{x_g},
\end{equation}
where the Wilson line $U(x_{\perp})$ is defined as
$U^n\left[-\infty,+\infty; x_{\perp}\right]$. At small-$x$ for a
large nucleus, this distribution can be evaluated using the
McLerran-Venugopalan model\footnote{To obtain this result, it was
assumed that the color charge densities in the nucleus obey a
Gaussian distribution with variance $\mu^2$. It was recently
argued that this assumption is inconsistent with the QCD
non-linear evolution~\cite{Dumitru:2010ak}, except for two-point
functions.
%FY
%However, this assumption is not necessary, in general the WW gluon distribution is obtained from a 4-point %function in the CGC.
}~\cite{McLerran:1993ni}
\begin{equation}
xG^{(1)}(x,k_{\perp })=\frac{S_{\perp }}{\pi ^{2}\alpha _{s}}\frac{%
N_{c}^{2}-1}{N_{c}}\int \frac{d^{2}r_{\perp }}{(2\pi )^{2}}\frac{%
e^{-ik_{\perp }\cdot r_{\perp }}}{r_{\perp }^{2}}\left( 1-e^{-\frac{r_{\perp
}^{2}Q_{s}^{2}}{4}}\right) \ ,
\label{mvmodel}
\end{equation}%
where $N_c=3$ is the number of colors, $S_\perp$ is the transverse area of the target nucleus, and $Q^2_{s}=\frac{g^2N_c}{4\pi}\ln\frac{1}{r_{\perp}^2\lambda^2}\int dx^{-} \mu^2(x^{-})$ is the gluon saturation
scale~\cite{Iancu:2003xm} with $\mu^2$ the color charge density in a large nuclei. We have cross checked this result by directly
calculating the gluon distribution function in Eq.~(\ref{g1}) following the
similar calculation for the quark in Ref.~\cite{Belitsky:2002sm,
Brodsky:2002ue}. The derivation of the WW gluon distribution from its operator definition is provided in Appendix~\ref{wwg}.

The second gluon distribution, the Fourier transform of the dipole
cross section, is defined in the fundamental
representation\footnote{The Fourier transform of the dipole cross
section in the adjoint representation is also commonly used, as it
enters single gluon production in $pA$ collisions
\cite{Kovner:2001vi,Kovchegov:2001sc,Marquet:2004xa}. In the
large-$N_c$ limit, it is related to the convolution of two
$xG^{(2)}$.}
\begin{eqnarray}
xG^{(2)}(x,k_{\perp }) &=&2\int \frac{d\xi ^{-}d\xi _{\perp }}{(2\pi
)^{3}P^{+}}e^{ixP^{+}\xi ^{-}-ik_{\perp }\cdot \xi _{\perp }}\langle P|\text{%
Tr}\left[F^{+i}(\xi ^{-},\xi _{\perp })\mathcal{U}^{[-]\dagger }F^{+i}(0)%
\mathcal{U}^{[+]}\right]|P\rangle \ ,  \label{g2}
\end{eqnarray}
where the gauge link
$\mathcal{U}_\xi^{[-]}=U^n\left[0,-\infty;0\right]U^n\left[-\infty,
\xi^{-}; \xi_{\perp}\right]$ stands for initial state
interactions. Thus, the dipole gluon distribution contains both
initial and final state interactions in the definition.

$\mathcal{U}^{[+]}$ and $\mathcal{U}^{[-]}$ are the gauge links
which appear in the quark distributions in the DIS and Drell-Yan
process, respectively. It is well-known that there is only final
state effect in the DIS, while there is only initial state
interaction in the Drell-Yan process. In addition, in processes
involving gluons and more complicated partonic structures, more
complex gauge links may appear, such as combinations of
$\mathcal{U}^{[+]}$ and $\mathcal{U}^{[-]}$~\cite{Bomhof:2006dp}.
We will see this in our following calculations especially in dijet
production in $pA$ collisions.

For the second gluon distribution $xG^{(2)}$ as shown in
Eq.~(\ref{g2}), the gauge link contribution can not be completely
eliminated. In other words, there is no number density
interpretation for this gluon distribution. This is also because
it contains both initial and final state interaction effects. Due
to the gauge link in this gluon distribution from $-\infty$ to
$+\infty$, naturally this gluon distribution can be related to the
color-dipole cross section evaluated from a dipole of size
$r_{\perp}$ scattering on the nucleus target, and has been
calculated in the CGC formalism,
\begin{equation}
xG^{(2)}(x,q_\perp)= \frac{q_{\perp }^{2}N_{c}}{2\pi^2 \alpha_s}%
S_{\perp }\int
\frac{d^2r_\perp}{(2\pi)^2}e^{-iq_\perp\cdot r_\perp}
\frac{1}{N_c}\left\langle\text{Tr}U(0)U^\dagger(r_\perp)\right\rangle_{x_g} .
\end{equation}
The derivation of this dipole gluon distribution from its operator definition is provided in Appendix~\ref{dipoleg}.

These two gluon distributions have been intensively investigated
in the last few years\footnote{There have been an observation that
these two UGDs can be related through a mathematical formula
$xG^{(2)}_g (x,q_\perp) \propto q_\perp^2\nabla^2_{q_\perp}
xG^{(1)}_g (x,q_\perp)$ where $xG^{(2)}_g (x,q_\perp)$ stands for
the gluon distribution in the adjoint representation which is
derived from a dipole formed by two gluons (e.g., see
ref.~\cite{Kharzeev:2003wz}). However, we believe that this
relation is just a mathematical observation without any physics
derivation. In addition, we find that it only works for MV model
which assumes the local gaussian approximation. This mathematical
relation is invalidated beyond the local gaussian approximation.
(e.g., see Appendix~\ref{wwg}.) From the above operator definition
of these two UGDs, we can see that they are two independent gluon
distributions. }. In particular, it was found that they
\begin{itemize}
\item have the same perturbative behavior. They both scale as $Q_s^2/q^2_\perp$ at
large transverse momentum $q_\perp\gg Q_s$;
\item however, they differ dramatically at small transverse momentum:
$G^{(1)}\sim \ln Q_s^2/q_\perp^2$ whereas $G^{(2)}\sim q_\perp^2$.
\end{itemize}
It will be very important to test these predictions by measuring
the quark-antiquark correlation in DIS process and direct photon
jet correlation in $pA$ collisions, since these processes can
directly probe these two gluon distributions separately.

The second gluon distribution ($xG^{(2)}$) depends on the dipole cross section,
which appears in various inclusive and semi-inclusive processes. For example,
the total cross section (or the structure functions) in DIS, the single inclusive hadron
production in DIS and $pA$ collisions, and the Drell-Yan lepton pair production in
$pA$ collisions, are all depending on this dipole gluon distribution. Tremendous
phenomenological analysis have been performed to constrain this gluon distribution
from the experimental data.

On the other hand, the first gluon distribution ($xG^{(1)}$) only appears
in few physical processes. Thus, we do not
have much constraints on its behavior. The only knowledge comes from model
calculations (i.e., the GBW model\cite{GolecBiernat:1998js} which provides a good description of all DIS data below
x = 0.01). Therefore, it is very crucial to carry out experimental observation of the
quark-antiquark jet correlation in DIS process in the planed Electron-Ion collider, which
shall provide very important information on this gluon distribution.

Two particle production in $pA$ collisions are found to depend on
both gluon distributions~\cite{Dominguez:2010xd}. In Table I, we
summarize the current status for the two gluon distributions
probed in high energy processes, where we find that the dipole
gluon distribution contributes to most of them, such as inclusive
DIS, semi-inclusive DIS(SIDIS)~\cite{Marquet:2009ca}, Drell-Yan
(DY) processes, single inclusive hadron production in $pA$
collisions, photon-jet correlations and dijet in $pA$ collisions,
whereas the WW gluon distribution only appears in the
quark-antiquark dijet correlation in DIS and dijet correlations in
$pA$ collisions. It is important to note that our derivations for the two basic processes, where the two distributions can be measured independently (dijet correlations in DIS and photon-jet correlation in $pA$ collisions), are exact for finite $N_c$. The large-$N_c$ limit is only necessary for more complicated processes where it allows us to write the new distributions as convolutions of the two basic ones.
\begin{table}
\begin{ruledtabular}
\begin{tabular}{|l|c|c|c|c|c|c|}
% \hline
& DIS and DY & SIDIS & hadron in $pA$ & photon-jet in $pA$ & Dijet in DIS & Dijet in $pA$  \\
\hline $G^{(1)}$ (WW)&  $\times$ & $\times$ &$\times$&$\times$&$\surd$&$\surd$\\
\hline $G^{(2)}$ (dipole)&  $\surd$ & $\surd$ &$\surd$&$\surd$&$\times$&$\surd$
 \end{tabular}
\end{ruledtabular}
\caption{The involvement of these two gluon distributions in high energy processes.}
\end{table}

In the following sections, we will carry out the detailed derivations for the two-particle
correlations in these processes. Quark-antiquark correlation in DIS process will be
calculated in Sec.II. Sec.III will be devoted to the direct photon jet correlation in
$pA$ collisions. We will derive the formalism for dijet correlation in
$pA$ collisions in Sec. IV. Summary and further discussions will be given
in Sec. V. In all these calculations, we will show the results from both transverse
momentum dependent approach and the CGC calculations and we will demonstrate
that they are consistent in the correlation limit.

\section{Dijet production in DIS}

\begin{figure}[tbp]
\begin{center}
\includegraphics[width=12cm]{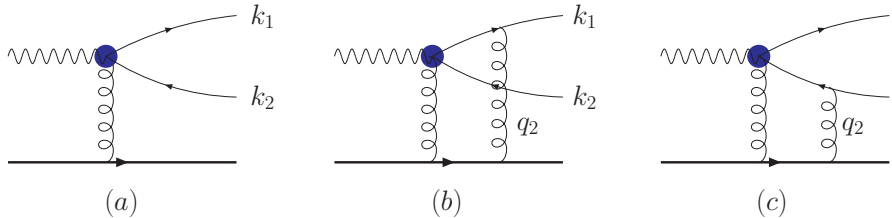} \\[0pt]
\end{center}
\caption[*]{Typical Feynman diagrams contributing to the quark-antiquark jet
correlation in deep inelastic scattering: (a) leading order, where the bubble represents
the gluon attachments to both quark lines; (b,c) two-gluon exchange
diagrams. }
\label{fig1}
\end{figure}

Despite the nice physical interpretation, it has been argued that the gluon
distribution in Eq.~(\ref{g1}) is not directly related to physical
observables in the CGC formalism. However, we will show that $xG^{(1)}$ can
be directly probed through the quark-antiquark jet correlation in DIS,
\begin{equation}
\gamma_T^{\ast }A\rightarrow q(k_{1})+\bar{q}(k_{2})+X\ ,  \label{dis}
\end{equation}%
where the incoming (virtual) photon carries momentum $k_{\gamma^*}$,
the target nucleus has momentum $P_A$, and the final state quark
and antiquark with momenta $k_1$ and $k_2$, respectively.
Again, we focus on the kinematic region with the correlation limit: $%
q_{\perp }=|\vec{k}_{1\perp }+\vec{k}_{2\perp }|\ll P_\perp$. The transverse
momenta are defined in the center of mass frame of the virtual photon $%
\gamma^*$ and the nucleus $A$. The calculations are performed for $Q^2$ in
the same order of $P_\perp^2$. As we
discussed in the above, we take the leading order contribution in the
correlation limit: $q_\perp\ll P_\perp$, and neglect all higher order corrections. We plot the
typical Feynman diagram for the process of (\ref{dis}) in Fig.~\ref{fig1}, where the
bubble in the partonic part represents the hard interaction vertex including
gluon attachments to both quark and antiquark lines. Fig.~\ref{fig1} (a) is the
leading Born diagram whose contributions can be associated with the hard partonic
cross section times the gluon distribution from Eq.~(\ref{g1})~\cite%
{qvy-short}. In high energy scattering with the nucleus target, additional
gluon attachments are important and we have to resum these contributions in
the large nuclear number limit. Figs.~1(b,c) represent the diagrams
contributing at two-gluon exchange order, where the second gluon can attach
to either the quark line or the antiquark line. By applying the power
counting method in the correlation limit ($q_\perp\ll P_\perp$), we can
simplify the scattering amplitudes with the Eikonal approximation~\cite%
{qvy-short}. For example, Fig.~\ref{fig1} (b) can be reduced to:
\begin{equation}
 \frac{g}{-q_{2}^{+}+i\epsilon} T^{b}\Gamma^{a} \ ,
\end{equation}
where $q_{2}$ is the gluon
momentum, $T^b$ is the $SU(3)$ color matrix in the fundamental
representation and $\Gamma^{a}$ represents the rest of the partonic
scattering amplitude with color indices for the two gluons $a$ and $b$.
Similarly, Fig.~\ref{fig1}(c) can be reduced to:
\begin{equation}
-\frac{g}{-q_{2}^{+}+i\epsilon} \Gamma^{a}T^{b}\ .
\end{equation}
The sum of these two diagrams will be ${g}/(-q_{2}^{+}+i\epsilon
)\left[T^b\Gamma^a-\Gamma^aT^b\right]$. Because of the unique color index in
$\Gamma _{a}$, we find the effective vertex as,
\begin{equation}
\mathrm{Fig.~\ref{fig1}(b,c)}\sim \frac{i}{-q_{2}^{+}+i\epsilon }%
(-ig)(-if_{bca})T^{c} \ ,
\end{equation}%
which corresponds to the first order expansion of the gauge link contribution in
the gluon distribution defined in Eq.~(\ref{g1}). For all high order
contributions, we can follow the procedure outlined in
Ref.~\cite{Belitsky:2002sm,{Bomhof:2006dp}} to derive
the gluon distribution.

%FY
In particular, we calculate the differential cross section
contributions from the diagrams of Fig.~\ref{fig1}, assuming the
generic coupling between the exchanged gluons and the nucleus
target. The contributions at given order can be reproduced by
the hard partonic cross section (given below) multiplying the TMD
gluon distribution defined as Eq.~(\ref{g1}) at the same
order from the similar diagrams. This method is particular
useful to identify the gluon distributions involved in the
hard scattering processes and will be applied throughout the
following calculations.

Of course, to build a rigorous TMD factorization theorem for this process, we
have to go beyond the diagrams shown in Fig.~\ref{fig1}, and
include the real gluon radiation contributions~\cite{Collins:1981uw,{Ji:2005nu}}.
These diagrams will introduce the large logarithms of
$\ell n (P_\perp^2/q_\perp^2)$, in addition to the small-$x$
logarithms $\ell n (1/x)$. The combination of both effects
has not yet been systematically studied in the literature.
We hope to address this issue in the future.
Moreover, there have been discussions on the power counting
method to factorize the gluon distribution from any
generic Feynman diagrams,
where one has to be extra cautions about the
``super-leading-power" contributions (see, for example,
Ref.~\cite{Collins:2008sg}).

From our analysis, we identified the gluon distribution
involved in the quark-antiquark jet correlation in DIS
process is the first gluon distribution at small-$x$.
We want to emphasize that this result is also the unique
consequence of non-abelian feature of QCD. For Abelian
theory, we can easily find that the final state interactions
between the quark and antiquark with the nucleus target
cancel out completely.
Therefore, there is no final state interaction effects
in the similar QED process.
%FY

Furthermore, the gauge link $\mathcal{U}^{[+]}$ in
Eq.~(\ref{g1fund}) can be viewed as the sum of all the final state
interactions between the nucleus target and the produced quark as
shown in Fig.~\ref{fig1} (b). In the meantime, the gauge link
$\mathcal{U}^{[+]\dagger}$ in Eq.~(\ref{g1fund}) takes care of the
final state interactions between the nucleus target and the
produced antiquark as illustrated in Fig.~\ref{fig1} (c).
Therefore, following Ref.~\cite{Bomhof:2006dp}, it is
straightforward to show that the Weizs\"acker-Williams gluon
distribution is the relevant gluon distribution in DIS dijet since
it correctly resums all the final state interactions.

\subsection{TMD-factorization approach to the DIS dijet production}

By putting in the hard partonic cross section $H_{\gamma^*g\to
q\bar q}$ and especially the correct gluon distribution, namely
the WW gluon distribution, which resums all the final state
interactions between the $q\bar{q}$ pair and the target nucleus,
we obtain the following transverse and longitudinal differential
cross sections for the quark-antiquark jet correlation in DIS
process
\begin{eqnarray}
\frac{d\sigma_{\textrm{TMD}}^{\gamma_T^{\ast }A\rightarrow q\bar{q}+X}}{d\mathcal{P.S.}}
&=&\delta (x_{\gamma ^{\ast }}-1) x_{g}G^{(1)}(x_{g},q_{\perp
})H_{\gamma_T^*g\to q\bar q}, \label{factdis}\\
\frac{d\sigma_{\textrm{TMD}}^{\gamma_L^{\ast }A\rightarrow q\bar{q}+X}}{d\mathcal{P.S.}}
&=&\delta (x_{\gamma ^{\ast }}-1) x_{g}G^{(1)}(x_{g},q_{\perp
})H_{\gamma_L^*g\to q\bar q} ,  \label{disdj}
\end{eqnarray}
where $x_g$ is the momentum fraction of hadron $A$ carried by the gluon and is determined
by the kinematics, $x_{\gamma^*}=z_q+z_{\bar q}$ with $z_q=z$ and $z_{\bar q}=1-z$
being the momentum fractions of the virtual photon carried by the quark and
antiquark, respectively. The phase space factor is defined as $d\mathcal{P.S.
}= dy_{1}dy_{2}d^2P_{\perp }d^{2}q_{\perp }$, and $y_1$ and $y_2$ are
rapidities of the two outgoing particles in the lab frame. In terms of the rapidities and the center of mass energy $\sqrt{s}$, one can find
\begin{equation}
z=\frac{|k_{1\perp}|e^{y_1}}{|k_{1\perp}|e^{y_1}+|k_{2\perp}|e^{y_2}}, \quad x_{\gamma^*}=\frac{|k_{1\perp}|e^{y_1}+|k_{2\perp}|e^{y_2}}{\sqrt{s}}, \quad x_{g}=\frac{|k_{1\perp}|e^{-y_1}+|k_{2\perp}|e^{-y_2}}{\sqrt{s}}.
\end{equation}
In addition, in the correlation limit, one has $|P_{\perp}|\simeq |k_{1\perp}| \simeq |k_{2\perp}| \gg |q_{\perp}|=|k_{1\perp}+k_{2\perp}|$. The leading order hard partonic
cross section reads
\begin{eqnarray}
H_{\gamma_T^*g\to q\bar q}&=&{\alpha_{s}\alpha _{em}e_{q}^{2}}\frac{\hat
s^2+Q^4}{(\hat s+Q^2)^4} \left(\frac{\hat{u}}{\hat{t}}+\frac{\hat{t}}{\hat{u}%
}\right) \\
H_{\gamma_L^*g\to q\bar q}&=&{\alpha_{s}\alpha _{em}e_{q}^{2}}\frac{8\hat{s}%
Q^2}{(\hat s+Q^2)^4}
\end{eqnarray}
with the usually defined partonic Mandelstam variables $\hat
s=(k_1+k_2)^2=P_\perp^2/(z(1-z))$, $\hat t=(k_2-k_{\gamma^*})^2=
-(P_\perp^2+\epsilon_f^2)/(1-z)$, and $\hat u=(k_1-k_{\gamma^*})^2=
-(P_\perp^2+\epsilon_f^2)/z$ with $\epsilon_f^2=z(1-z)Q^2$ and $z=z_q$.
%FY

Finally, in the correlation limit, one obtains the differential total cross section as follows:
\begin{eqnarray}
\frac{\text{d}\sigma_{\textrm{tot}}^{\gamma^{*} A\to q\bar{q}X}}{\text{d}y_1 \text{d}y_2 \text{d}^2P_{\perp}\text{d}^2 q_{\perp}}&=&\delta(x_{\gamma^*}-1)\frac{z(1-z)}{(P^{2}_{\perp}+\epsilon_f^2)^4}\left[\left(z^2+(1-z)^2
\right)(P^{4}_{\perp}+\epsilon_f^4)+8z(1-z)P^{2}_{\perp} \epsilon_f^2\right]\notag \\
&&\times\frac{
S_{\perp}N_c \alpha_{em} e^2_q}{4\pi^4}\int \text{d}^2 r_{\perp}e^{-iq_{\perp}\cdot r_{\perp}}\frac{1}{
r_{\perp}^2}\left[1-\exp \left(-\frac{1}{4}r^{2}_{\perp}Q_s^2\right)\right]\ ,
\end{eqnarray}
where $\sigma_{\rm tot}$ is defined as $\sigma_{\rm tot}=\sigma_T+\sigma_L$
and we have substituted the CGC result for the WW gluon distribution in Eq.~(\ref{mvmodel}).
By taking $Q^2=0$, we can extend the above result to the case of dijet
production in real photon scattering on nuclei. The longitudinal
contribution vanishes and the total cross section only contains the
transverse part. Therefore, we obtain
\begin{eqnarray}
\frac{\text{d}\sigma^{\gamma A\to q\bar{q}X}}{\text{d}y_1 \text{d}y_2 \text{d%
}^2P_{\perp}\text{d}^2 q_{\perp}}&=&\delta(x_{\gamma}-1) \frac{%
S_{\perp}N_c \alpha_{em} e^2_q}{4\pi^4P^{4}_{\perp}}z(1-z)\left[z^2+(1-z)^2%
\right]  \notag \\
&&\times\int \text{d}^2 r_{\perp}e^{-iq_{\perp}\cdot r_{\perp}}\frac{1}{%
r_{\perp}^2}\left[1-\exp \left(-\frac{1}{4}r^{2}_{\perp}Q_s^2\right)\right].
\end{eqnarray}
%FY
For the real photon case, there will be resolved photon contributions which
should be taken care separately following that in the dijet production
in $pA$ collisions discussed in Sec. IV below.
%FY

\subsection{CGC approach to the DIS dijet production}\label{discgc}

\begin{figure}[tbp]
\begin{center}
\includegraphics[width=9cm]{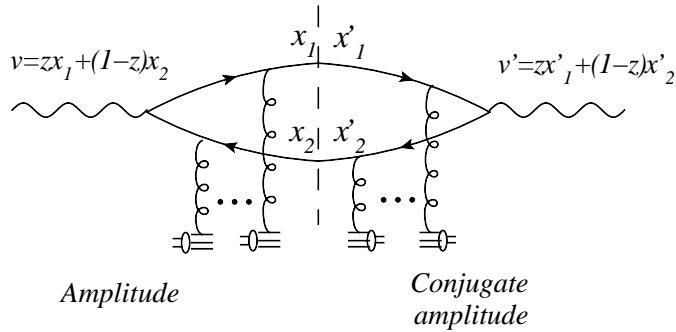} \\[0pt]
\end{center}
\caption[*]{Typical diagrams contributing to the cross section in
the deep inelastic process.} \label{discgc2}
\end{figure}

The quark-antiquark jet cross section can also be calculated in
the CGC formalism. In this setup the photon splits into a
quark-antiquark pair which subsequently undergoes multiple
interactions with the nucleus (see Fig. \ref{discgc2}). Previous
calculations performed under this framework
\cite{Kovner:2001vi,Gelis:2002nn} have focused mainly on the total
cross section or single inclusive gluon production, which are
calculations involving a different color structure than the
process we are interested in. Here we calculate the cross section
in the most general case and then we show how the factorization
formula is recovered in the correlation limit.

At the amplitude level the process can be divided into two parts:
the splitting wave function of the incoming photon and the
multiple scattering factor. It is convenient to write these
quantities in transverse coordinate space since in this basis, and
in the eikonal approximation, the multiple interaction factor is
diagonal.
%FY
To be consistent with previous CGC calculations in the literature, we choose
a frame that the photon is moving along the $+\hat z$ direction whereas the nuclear target
in the $-\hat z$ direction. When we compare the results to those obtained in the
last subsection, we have to keep in mind this difference. However, we note that
the differential cross section does not depend on the frame.
%FY
For a right-moving photon  with longitudinal momentum $p^+$, no
transverse momentum, and virtuality $Q^2$, the splitting wave
function in transverse coordinate space takes the form,
\begin{align}
\psi^{T\,\lambda}_{\alpha\beta}(p^+,z,r)&=2\pi\sqrt{\frac{2}{p^+}}\begin{cases}i\epsilon_fK_1(\epsilon_f|r|)\tfrac{r\cdot\epsilon^{(1)}_\perp}{|r|}[\delta_{\alpha+}\delta_{\beta+}(1-z)+\delta_{\alpha-}\delta_{\beta-}z], & \lambda=1,\\ i\epsilon_fK_1(\epsilon_f|r|)\tfrac{r\cdot\epsilon^{(2)}_\perp}{|r|}[\delta_{\alpha-}\delta_{\beta-}(1-z)+\delta_{\alpha+}\delta_{\beta+}z], & \lambda=2,\end{cases}\\
\psi^L_{\alpha\beta}(p^+,z,r)&=2\pi\sqrt{\frac{4}{p^+}}z(1-z)QK_0(\epsilon_f|r|)\delta_{\alpha\beta}.
\end{align}
where again $z$ is the momentum fraction of the photon carried by the quark,
$\lambda$ is the photon polarization, $\alpha$ and $\beta$ are the quark and
antiquark helicities, $r$ the transverse separation of the pair, $\epsilon_f^2=z(1-z)Q^2$,
and the quarks are assumed to be massless. The heavy quark case will be considered in the next subsection.

The multiple scattering factor is expressed in terms of Wilson lines in the fundamental
representation. It can be shown~\cite{Gelis:2002nn} that this interaction term takes the
form $\left[U^\dagger(x_2)U(x_1)-1\right]_{ji}$ where $x_1$ and $x_2$ are the
transverse positions of the quark and the antiquark, $i$ and $j$ are their color
indices, and the Wilson line is given in terms of the background field by
\begin{equation}
U(x)=\mathcal{P}\exp\left\{ig_S\int_{-\infty}^{+\infty} \text{d}x^+\,T^cA_c^-(x^+,x)\right\} \ .
\end{equation}
%FY
%Here we are using the light-cone gauge of the projectile $A^+=0$ and, for
%a left-moving nucleus, the only non-zero component of the gauge field is $A^-$
%(see \cite{Gelis:2005pt}).
%FY
The gauge field is directly related to the color charge density of the nucleus
which will be averaged over the nuclear wave function at the level of the cross
section. The way the color indices are contracted in the scattering factor is due
to the fact that the pair is initially in a singlet state but no assumptions are made
about the final state. The color indices $i$ and $j$ will be summed over
independently also at the cross section level.

With the pieces described above we can write down an explicit formula for the
differential cross section for dijet production. After averaging over the photon's
polarization and summing over the quark and antiquark helicities and colors we obtain,
\begin{eqnarray}
\frac{d\sigma ^{\gamma_{T,L}^{\ast }A\rightarrow q\bar{q}X}}{d^3k_1d^3k_2}
&=&N_{c}\alpha _{em}e_{q}^{2}\delta(p^+-k_1^+-k_2^+) \int
\frac{\text{d}^{2}x_1}{(2\pi)^{2}}\frac{\text{d}^{2}x_1^{\prime }}{(2\pi )^{2}}
\frac{\text{d}^{2}x_2}{(2\pi)^{2}}\frac{\text{d}^{2}x_2^{\prime }}{(2\pi )^{2}} \notag \\
&&\times e^{-ik_{1\perp }\cdot(x_1-x_1^{\prime })} e^{-ik_{2\perp }\cdot (x_2-x_2^{\prime })}
\sum_{\lambda\alpha\beta} \psi_{\alpha\beta}^{T, L \lambda}(x_1-x_2)
\psi_{\alpha\beta}^{T, L\lambda*}(x_1^{\prime }-x_2^{\prime })  \notag \\
&&\times \left[1+S^{(4)}_{x_g}(x_1,x_2;x_2^{\prime },x_1^{\prime})
-S^{(2)}_{x_g}(x_1,x_2)-S^{(2)}_{x_g}(x_2^{\prime },x_1^{\prime })\right] \ ,\label{xsdis}
\end{eqnarray}
where the two- and four-point functions are defined as
\begin{eqnarray}
&&S_{x_g}^{(2)}(x_1,x_2)=\frac{1}{N_c}\left\langle\text{Tr}U(x_1)U^\dagger(x_2)\right\rangle_{x_g}\ , \\
&&S_{x_g}^{(4)}(x_1,x_2;x_2^{\prime},x_1^{\prime})=\frac{1}{N_c}\left\langle\text{Tr}U(x_1)U^\dagger(x_1^{\prime})
U(x_2^{\prime})U^\dagger(x_2)\right\rangle_{x_g}\ .
\end{eqnarray}
The notation $\langle\dots\rangle_{x_g}$ is used for the CGC average of the color
charges over the nuclear wave function where $x_g$ is the smallest fraction of
longitudinal momentum probed, and is determined by the kinematics.

Notice that the transverse coordinates of the quark and antiquark in the amplitude
(unprimed coordinates) are different from the coordinates in the complex conjugate
amplitude (primed coordinates) since the two final momenta are not integrated over.
This is a very important feature of our calculation that, to our knowledge, does not
appear in previous CGC calculations of DIS in nuclei. It allows for a different color
structure and in particular it is responsible for the appearance of the 4-point function
$S^{(4)}_{x_g}$ which cannot be expressed in terms of 2-point functions, even in the
large $N_c$ limit (see Appendix~\ref{4point} for an explicit evaluation of the medium average).

In order to compare with the TMD-factorization result discussed in the previous section,
we need to consider the relevant kinematic region, in particular in the correlation limit of
Eq. (\ref{xsdis}). For convenience, we introduce the transverse coordinate variables:
$u=x_1-x_2$ and $v=zx_1+(1-z)x_2$, and similarly for the primed coordinates.
The respective conjugate momenta are $\tilde P_\perp=(1-z)k_{1\perp}-zk_{2\perp}\approx P_\perp$
and $q_\perp$, and therefore the correlation limit can be taken by assuming $u$ and $u'$
are small and then expanding the integrand with respect to these two variables before
performing the Fourier transform.

Let us focus on the multiple scattering factor first. By using the following
identities,
\begin{eqnarray}
S^{(4)}_{x_g}(x_1,x_2;v',v')&=&S^{(2)}_{x_g}(x_1,x_2)\ ,\\ S^{(4)}_{x_g}(v,v;x'_2,x'_1)&=&S^{(2)}_{x_g}(x'_2,x'_1)\ ,
\end{eqnarray}
it is easy to see that terms from the expansion of $S^{(4)}_{x_g}$ cancel the other
terms in (\ref{xsdis}). After applying  $$U^\dagger(v)\left(\partial_iU(v)\right)=-\left(\partial_iU^\dagger(v)\right)U(v)\ ,$$
we can show that the lowest order contribution in $u$ and $u'$ to the scattering
factor can be written as
\begin{equation}
-u_iu'_j\frac{1}{N_c}\langle\text{Tr}\left[\partial_iU(v)\right]U^\dagger(v')\left[\partial_jU(v')\right]
U^\dagger(v)\rangle_{x_g}\ .
\end{equation}
Taking into account the path ordering of the Wilson lines, we have the
following formula for their derivatives,
\begin{equation}
\partial_iU(v)=ig_S\int_{-\infty}^\infty \text{d}v^+\,U[-\infty,v^+;v]\,\left(\partial_iA^-(v^+,v)\right)\,U[v^+,\infty;v],
\end{equation}
where $U[a,b;x]=\mathcal{P}\exp\{ig_S\int_a^b\text{d}x^+\,T^cA_c^-(x^+,x)\}$.
%FY
We notice that $\left(\partial_iA^-(v^+,v)\right)$ is part of the
gauge invariant field strength tensor $F^{i-}(\vec{v})$\footnote{The other part
of the the field strength tensor shall come from the transverse component
of the Wilson lines as the gauge invariance of QCD requires. When the $A^+=0$ gauge is used the only non-zero component of the gauge field is $A^-$ \cite{Gelis:2005pt} and the transverse parts drop out of the equations, giving a simpler form of the equations.}.
%FY
Therefore, the above correlator can be written in terms of
gauge invariant matrix element,
\begin{equation}
-\langle\text{Tr}\left[\partial_iU(v)\right]U^\dagger(v')\left[\partial_jU(v')\right]U^\dagger(v)
\rangle_{x_g}=g_S^2\int_{-\infty}^\infty\text{d}v^+\text{d}v^{\prime+}
\left\langle\text{Tr}\left[F^{i-}(\vec{v})\mathcal{U}^{[+]\dagger}F^{j-}(\vec{v}')
\mathcal{U}^{[+]}\right]\right\rangle_{x_g} \ .
\end{equation}
%FY
%where we have used the fact that, in the gauge we are using, the only non-zero component of the gauge %field is $A^-$.
Performing the $u$ and $u'$ integration in (\ref{xsdis}) after the
expansion of the multiple scattering term, we find an explicit
formula for the differential cross section in the desired kinematic region,
\begin{eqnarray}
\frac{d\sigma ^{\gamma_T ^{\ast }A\rightarrow q\bar{q}X}}{d\mathcal{P.S.}}
&=&\alpha_{em}e_{q}^{2}\alpha_s\delta
\left(x_{\gamma^*}-1\right)z(1-z)\left(z^{2}+(1-z)^{2}\right) \frac{%
P_\perp^4+\epsilon_f^4}{(P_{\perp }^{2}+\epsilon_f^2)^4}  \notag \\
&&\times (16\pi^3)\int \frac{d^{3}v d^3v^{\prime }}{(2\pi )^{6}}%
e^{-iq_{\perp }\cdot (v-v^{\prime })}2\left\langle\text{Tr}\left[F^{i-}({v})%
\mathcal{U}^{[+]\dagger}F^{i-}({v}^{\prime}) \mathcal{U}^{[+]}\right]\right%
\rangle_{x_g} \ , \label{dipoledis} \\
\frac{d\sigma ^{\gamma_L ^{\ast }A\rightarrow q\bar{q}X}}{d\mathcal{P.S.}}
&=&\alpha_{em}e_{q}^{2}\alpha_s\delta
\left(x_{\gamma^*}-1\right)z^2(1-z)^2 \frac{8P_\perp^2\epsilon_f^2}{(P_{\perp }^{2}+\epsilon_f^2)^4}  \notag \\
&&\times (16\pi^3)\int \frac{d^{3}v d^3v^{\prime }}{(2\pi )^{6}}%
e^{-iq_{\perp }\cdot (v-v^{\prime })}2\left\langle\text{Tr}\left[F^{i-}({v})%
\mathcal{U}^{[+]\dagger}F^{i-}({v}^{\prime}) \mathcal{U}^{[+]}\right]\right%
\rangle_{x_g} \label{dipoledisl} \ .
\end{eqnarray}
These results are to be compared to the factorized results in Eq. (\ref{factdis},\ref{disdj}).
The hard cross section factor in (\ref{factdis}) is recovered by noticing that in the
kinematic region we are considering the Mandelstam variables are given by
$\hat s=P_\perp^2/(z(1-z))$, $\hat t=-(P_\perp^2+\epsilon_f^2)/(1-z)$, and
$\hat u=-(P_\perp^2+\epsilon_f^2)/z$. To recover the gluon distribution function
as written in Eq. (\ref{g1fund}) it is necessary to account for the different normalizations
used to calculate the average of Wilson lines above. In Eq. (\ref{g1fund}) the average
is calculated with a definite momentum (and therefore translational invariant)
hadronic state $|P\rangle$ which is relativistically normalized to
$\langle P'|P\rangle=(2\pi)^32P^+\delta(P^+-P^{\prime+})\delta^{(2)}(P_\perp-P'_\perp)$,
while the average in Eqs. (\ref{dipoledis}) and (\ref{dipoledisl}) is taken over the CGC wave function
and is normalized such that $\langle1\rangle_{x_g}=1$. Using translational
invariance Eq. (\ref{g1fund}) can be written as
\begin{eqnarray}
xG^{(1)}(x,k_\perp)&=&\frac{4}{\langle P|P\rangle}\int \frac{d\xi_1^{-}d^2\xi _{1\perp }d\xi_2^{-}d^2\xi _{2\perp }}{(2\pi )^{3}}e^{ixP^{+}(\xi_1^{-}-\xi_2^-)-ik_{\perp }\cdot(\xi _{1\perp }-\xi_{2\perp})}  \notag \\
&&\times \langle P|\text{Tr}\left[F^{+i}(\xi_1^{-},\xi_{1\perp })\mathcal{U}^{[+]\dagger }F^{+i}(\xi_2^{-},\xi_{2\perp })\mathcal{U}^{[+]}\right]|P\rangle \ .
\end{eqnarray}
It is easy to see that the discrepancy between normalizations
is accounted for by the replacement
$\frac{\langle P|\dots|P\rangle}{\langle P|P\rangle}\to\langle\dots\rangle_{x_g}$,
giving complete agreement between the CGC approach and the factorized form in the small-$x$ region.

In the end of this subsection, we would like to compare the dijet production process in DIS to the inclusive and semi-inclusive DIS. As shown above, we derive that the dijet production cross section in DIS is proportional to the WW gluon distribution in the correlation limit. On the other hand, it is well-known that inclusive and semi-inclusive DIS involves the dipole cross section instead~\cite{Marquet:2009ca}, which can be related to the second gluon distribution. This might look confusing at first sight, so let us take a closer look at Eq.~(\ref{xsdis}). If one integrates over one of the outgoing momenta, say $k_1$, one can easily see that the corresponding coordinates in the amplitude and conjugate amplitude are identified ($x_1=x_1^{\prime}$) and, therefore, the four-point function $S_{x_g}^{(4)}(x_1,x_2;x_2^{\prime},x_1^{\prime})$ collapses to a two-point function $S^{(2)}_{x_g}(x_2,x_2^{\prime})$. As a result, The SIDIS and inclusive DIS cross section only depend on two-point functions, thus they only involve the dipole gluon distribution. Now we can see the unique feature of the dijet production process in DIS. By keeping the momenta of the quark and antiquark unintegrated, we can keep the full color structure of the four-point function which eventually leads to the WW gluon distribution in the correlation limit. Therefore, measuring the dijet production cross sections or dihadron correlations in DIS at future experimental facilities like EIC or LHeC would give us a first direct and unique opportunity to probe and understand the Weizs\"acker-Williams gluon distribution.

\subsection{Heavy quark production in DIS dijet}

In order to expand our calculation and include the possibility of
charm and bottom production, we now consider the finite quark mass
case. From the TMD point of view, having massive quarks modifies
the hard cross sections while the parton distributions remain the
same. The new leading order hard partonic cross sections read
\begin{eqnarray}
H_{\gamma_T^*g\to q\bar q}&=&{\alpha_{s}\alpha _{em}e_{q}^{2}}z^2(1-z)^2\left[\frac{P_{\perp}^4+\epsilon^{\prime 4}_f}{(P_{\perp}^2+\epsilon^{\prime 2}_f)^4} \left(\frac{\tilde{u}}{\tilde{t}}+\frac{\tilde{t}}{\tilde{u}
}\right)+\frac{2m_q^2 P_{\perp}^2}{z(1-z)(P_{\perp}^2+\epsilon^{\prime 2}_f)^4} \right],\\
H_{\gamma_L^*g\to q\bar q}&=&{\alpha_{s}\alpha _{em}e_{q}^{2}}\frac{8
Q^2}{(\tilde s+Q^2)^4}\left(\tilde s - \frac{m_q^2}{z(1-z)}\right),
\end{eqnarray}
where $\tilde s=(k_1+k_2)^2=(P_\perp^2+m^2_q)/(z(1-z))$, $\tilde t=(k_2-k_{\gamma^*})^2-m_q^2=-(P_\perp^2+\epsilon_f^{\prime 2})/(1-z)$,
and $\tilde u=(k_1-k_{\gamma^*})^2-m_q^2=-(P_\perp^2+\epsilon_f^{\prime 2})/z$ with
$\epsilon^{\prime 2}_f=z(1-z)Q^2+m^2_q$ and $z=z_q$.

In terms of the CGC approach, one needs to modify the dipole splitting wave functions as follows:
\begin{align}
\psi^{T\,\lambda}_{\alpha\beta}(p^+,z,r)&=2\pi\sqrt{\frac{2}{p^+}}\begin{cases}i\epsilon_f^{\prime}K_1(\epsilon_f^{\prime}|r|)\tfrac{r\cdot\epsilon^{(1)}_\perp}{|r|}[\delta_{\alpha+}\delta_{\beta+}(1-z)+\delta_{\alpha-}\delta_{\beta-}z]\\
\quad+\delta_{\alpha -}\delta_{\beta +}m_qK_0(\epsilon_f^{\prime}|r|), & \lambda=1,\\
i\epsilon_f^{\prime}K_1(\epsilon_f^{\prime}|r|)\tfrac{r\cdot\epsilon^{(2)}_\perp}{|r|}[\delta_{\alpha-}\delta_{\beta-}(1-z)+\delta_{\alpha+}\delta_{\beta+}z]\\
\quad+\delta_{\alpha +}\delta_{\beta -}m_qK_0(\epsilon_f^{\prime}|r|), & \lambda=2,\end{cases}\\
\psi^L_{\alpha\beta}(p^+,z,r)&=2\pi\sqrt{\frac{4}{p^+}}z(1-z)QK_0(\epsilon_f^{\prime}|r|)\delta_{\alpha\beta}\;.
\end{align}
Following the same procedure, it is easy to show that again both approaches agree in the correlation limit for heavy quark production. By setting $Q^2=0$, one can get the results for the heavy quark production in real photon-nucleus scattering.

\section{Direct-photon jet in $pA$ collisions}

Now let us turn our attention to the second gluon distribution. In this context, the simplest process where we can access this distribution is the direct photon-quark jet correlation in $pA$ collisions,
\begin{equation}
pA\rightarrow \gamma (k_{1})+q(k_{2})+X\ ,  \label{dy}
\end{equation}
where the incoming quark carries momentum $p$, and nucleus target
with momentum $P_A$, and outgoing photon and quark with momenta $k_1$
and $k_2$, respectively.
The analysis of this process follows that for the quark-antiquark
jet correlation in DIS process in the previous section.
\begin{figure}[tbp]
\begin{center}
\includegraphics[width=12cm]{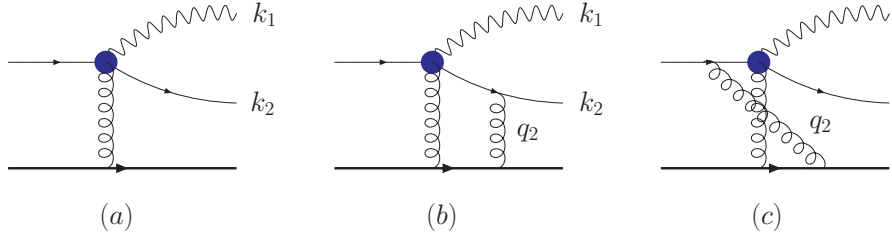}%\hfill
\end{center}
\caption[*]{Same as Fig.~(\ref{fig1}) for direct photon-jet correlation
in $pA$ collisions.}
\label{fig2}
\end{figure}
We plot the relevant diagrams in Fig.~\ref{fig2}(a,b,c), again for
the leading one gluon exchange and two gluon exchanges. Similarly,
the two gluon exchange contributions can be summarized as\footnote{There is a misprint, which we have corrected below in the eq.(\ref{col2}), in the Eq.(11) of the short summary \cite{Dominguez:2010xd} of this paper.)}
\begin{equation}
\mathrm{Fig.~\ref{fig2}(b,c)}\sim (-ig)\left(
\frac{i}{-q_{2}^{+}+i\epsilon }T^{b}\Gamma ^{a}+\frac{i}{q_{2}^{+}+i\epsilon }\Gamma ^{a}T^{b}\right) \ ,  \label{col2}
\end{equation}
where the plus sign comes from the fact that the second gluon attaches to
the quark line in the initial and final states. Since there is no color
structure corresponding to Eq.~(\ref{col2}), we can only express it in the
fundamental representation. Following Ref.~\cite{Bomhof:2006dp}, we find
that the gluon distribution in this process can be written as
\begin{eqnarray}
xG^{(2)}(x,k_{\perp }) &=&2\int \frac{d\xi ^{-}d\xi _{\perp }}{(2\pi
)^{3}P^{+}}e^{ixP^{+}\xi ^{-}-ik_{\perp }\cdot \xi _{\perp }}\langle P|\text{Tr}\left[F^{+i}(\xi ^{-},\xi _{\perp })\mathcal{U}^{[-]\dagger }F^{+i}(0)%
\mathcal{U}^{[+]}\right]|P\rangle \ ,  \label{g20}
\end{eqnarray}
where the gauge link $\mathcal{U}_\xi^{[-]}=U^n\left[0,-\infty;0\right]U^n
\left[-\infty, \xi^{-}; \xi_{\perp}\right]$ resums the initial state interactions between the incoming quark and the target nucleus. On the other hand, the gauge link $\mathcal{U}^{[+]}$ represents the final state interactions between the outgoing quark and the target nucleus. This gluon distribution can
also be calculated in the CGC formalism where it is found to be
\begin{equation}
xG^{(2)}(x,q_\perp)\simeq \frac{q_{\perp }^{2}N_{c}}{2\pi^2 \alpha_s}
S_{\perp }F_{x_{g}}(q_{\perp }),\label{g2f}
\end{equation}
with the normalized unintegrated gluon distribution $F_{x_g}(q_\perp)=\int
\frac{d^2r_\perp}{(2\pi)^2}e^{-iq_\perp\cdot r_\perp}
S_{x_g}^{(2)}(0,r_\perp)$. Therefore, by plugging in the appropriate gluon distribution, namely the dipole gluon distribution, which resums both the initial and final state interactions, one can write the differential cross section of (\ref{dy}) as\footnote{Here we assume that one can employ the collinear factorization for the integrated quark density or gluon density inside the dilute proton at large $x_p$, although the proof of this assumption is omitted throughout this paper. We will leave this study for future work. }
\begin{equation}
\frac{d\sigma^{\left( pA\rightarrow \gamma q+X\right)} }{d\mathcal{P.S.}}
=\sum_{f}x_{p}q_f(x_{p})x_gG^{(2)}(x_g,q_{\perp })H_{qg\to\gamma q}\ ,\label{factphoton}
\end{equation}%
where $x_1$ is the momentum fraction of the projectile nucleon carried by
the quark, $q_f(x_1)$ is the integrated quark distribution.
Because we are taking large nuclear number limit, the intrinsic
transverse momentum associated with it can be neglected compared to that
from the gluon distribution of nucleus. The hard partonic cross section is given by
\begin{equation}
H_{qg\to \gamma q}=\frac{\alpha_s\alpha_{em} e_q^2}{N_c\hat s^2}\left(-\frac{%
\hat s}{\hat u}-\frac{\hat u}{\hat s}\right).\label{hardph}
\end{equation}
Inserting Eqs. (\ref{g2f}) and (\ref{hardph}) in Eq. (\ref{factphoton}), one gets
\begin{equation}
\frac{d\sigma^{\left( pA\rightarrow \gamma q+X\right)} }{d\mathcal{P.S.}}
=\sum_{f}x_{p}q_{f}(x_{p})\frac{\alpha _{em}e_{f}^{2}}{2\pi ^{2}}%
S_{\perp }q_{\perp }^{2}F_{x_{g}}^{g}(q_{\perp }^{2})\ \frac{\left[ 1+\left(
1-z\right) ^{2}\right] z^{2}\left( 1-z\right) }{P_{\perp }^{4}}\ ,
\end{equation}
where we have expressed the Mandelstam variables in terms of $P_\perp$ and $z$:
$\hat{s}=(k_1+k_2)^2=\frac{P_{\perp }^{2}}{z\left( 1-z\right) }$, $\hat{u}=(k_1-p)^2=-\frac{%
P_{\perp }^{2}}{z}$ and $\hat{t}=(k_2-p)^2=-\frac{P_{\perp
}^{2}}{1-z}$. The momentum fraction of the incoming quark $p$
carried by the outgoing photon $z$ is defined as
\begin{equation}
z=\frac{|k_{1\perp}|e^{y_1}}{|k_{1\perp}|e^{y_1}+|k_{2\perp}|e^{y_2}},
\end{equation}
where $y_1$ and $y_2$ are rapidities of the photon and outgoing quark
in the Lab frame.

The current running RHIC and LHC experiments
shall provide us some information on the dipole gluon distribution
by measuring direct photon-quark jet correlation in $pA$
collisions.

\subsection{CGC approach to the direct photon-jet production in $pA$
collisions}

This process was already considered in the CGC framework in
\cite{Gelis:2002ki} where the calculation was performed entirely
in momentum space. In order to compare with the result from the
previous section and illustrate why a different distribution
should be used, we will derive the corresponding cross section
following the same procedure as the previous section by showing
the splitting wave function and the multiple scattering factor in
transverse coordinate space. Our result is consistent with
\cite{Gelis:2002ki}.

Let us consider the partonic level process $q\to q\gamma$. For a
right-moving massless quark, with initial longitudinal momentum
$p^+$ and no transverse momentum, the splitting wave function in
transverse coordinate space is given by
\begin{equation}
\psi^\lambda_{\alpha\beta}(p^+,k_1^+,r)=2\pi i\sqrt{\frac{2}{k_1^+}}\begin{cases}\frac{r\cdot\epsilon^{(1)}_\perp}{r^2}(\delta_{\alpha-}\delta_{\beta-}+(1-z)\delta_{\alpha+}\delta_{\beta+}), & \lambda=1,\\ \frac{r\cdot\epsilon^{(2)}_\perp}{r^2}(\delta_{\alpha+}\delta_{\beta+}+(1-z)\delta_{\alpha-}\delta_{\beta-}), & \lambda=2.\end{cases}\label{wvfunction} \ ,
\end{equation}
where again $\lambda$ is the photon polarization, $\alpha,\beta$ are helicities for the
incoming and outgoing quarks, and $z$ is the momentum fraction of the incoming quark
carried by the photon.
To account for the multiple scatterings in this process we have to consider interactions both before and after the splitting. If the transverse coordinates of the quark and photon in the final state are $b$ and $x$ respectively, then the multiple scattering factor in the amplitude takes the form $U(b)-U(zx+(1-z)b)$.

\begin{figure}[tbp]
\begin{center}
\includegraphics[width=9cm]{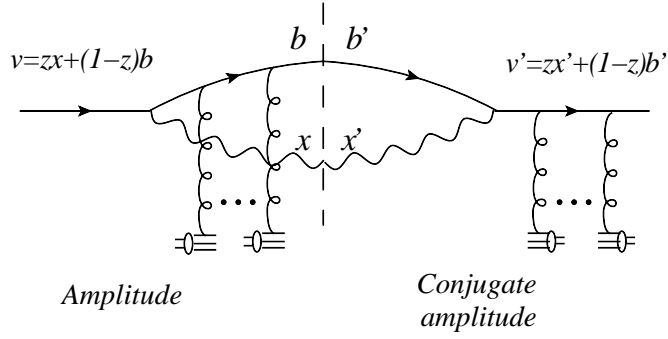} \\[0pt]
\end{center}
\caption[*]{Interactions before and after the splitting have to be taken into account for both amplitude and conjugate amplitude. Here is a typical diagram representing the third interaction term in Eq. (\ref{partonicphoton}).}
\label{photoncgc}
\end{figure}

After summing over final polarization, helicity and color, and averaging over initial helicity and color, we find the following expression for the partonic level cross section (see Fig. \ref{photoncgc}).
\begin{eqnarray}
\frac{d\sigma ^{qA\rightarrow q\gamma X}}{d^3k_1d^3k_2}&=&\alpha _{em}e_{q}^{2}\delta(p^+-k_1^+-k_2^+) \int \frac{\text{d}^{2}x}{(2\pi)^{2}}\frac{\text{d}^{2}x^{\prime }}{(2\pi )^{2}}\frac{\text{d}^{2}b}{(2\pi)^{2}}\frac{\text{d}^{2}b^{\prime }}{(2\pi )^{2}}  \notag \\
&&\times e^{-ik_{1\perp }\cdot(x-x^{\prime })}e^{-ik_{2\perp }\cdot (b-b^{\prime })} \sum_{\lambda\alpha\beta} \psi^{\lambda\ast}_{\alpha\beta}(x'-b')\psi^\lambda_{\alpha\beta}(x-b)  \notag \\
&&\times \left[S^{(2)}_{x_g}(b,b')+S^{(2)}_{x_g}(zx+(1-z)b,zx'+(1-z)b')\right.\notag \\
&&\left.-S^{(2)}_{x_g}(b,zx'+(1-z)b')-S^{(2)}_{x_g}(zx+(1-z)b,b')\right].\label{partonicphoton}
\end{eqnarray}

Notice that the color structure is simpler than in the DIS case. There is no four-point function and all the terms in the multiple scattering factor can be expressed in terms of the color dipole cross section $S^{(2)}_{x_g}$. By changing the variables on each of the terms of the scattering factor to $u=x-b$ and either $v=b$ or $v=zx+(1-z)b$ , and similarly for the primed coordinates, the cross section above can be written as
\begin{eqnarray}
\frac{d\sigma ^{qA\rightarrow q\gamma X}}{d^3k_1d^3k_2}&=&\alpha _{em}e_{q}^{2}\delta(p^+-k_1^+-k_2^+) \int \frac{\text{d}^{2}u}{(2\pi)^{2}}\frac{\text{d}^{2}u^{\prime }}{(2\pi )^{2}}\frac{\text{d}^{2}v}{(2\pi)^{2}}\frac{\text{d}^{2}v^{\prime }}{(2\pi )^{2}}  \notag \\
&&\times e^{-iq_{\perp}\cdot(v-v^{\prime })}S^{(2)}_{x_g}(v,v') \sum_{\lambda\alpha\beta} \psi^{\lambda\ast}_{\alpha\beta}(u')\psi^\lambda_{\alpha\beta}(u)  \notag \\
&&\times \left[e^{-iu\cdot(\tilde{P}_\perp+zq_\perp)}e^{iu'\cdot(\tilde{P}_\perp+zq_\perp)}+e^{-iu\cdot\tilde{P}_\perp}e^{iu'\cdot\tilde{P}_\perp}\right.\notag \\
&&\left.-e^{-iu\cdot(\tilde{P}_\perp+zq_\perp)}e^{iu'\cdot\tilde{P}_\perp}-
e^{-iu\cdot\tilde{P}_\perp}e^{iu'\cdot(\tilde{P}_\perp+zq_\perp)}\right]\ ,\label{xsphoton}
\end{eqnarray}
where $\tilde P_\perp=(1-z)k_{1\perp}-zk_{2\perp}\approx P_\perp$.

From the above expression it is easy to see that performing the $u$ and $u'$ integrations reduces to taking the Fourier transform of the splitting wave function with different values of the momentum variable for each term. Clearly, the Fourier transform of the dipole cross section factors out giving the gluon distribution we found from the TMD-factorized form. Using collinear approximation for the proton projectile we find our final result for the cross section of the desired process.
\begin{align}
\frac{d\sigma^{pA\rightarrow \gamma q+X} }{d\mathcal{P.S.}}
=&\;\sum_{f}x_{p}q_{f}(x_{p})\alpha _{em}e_{f}^{2}N_c\left[1+(1-z)^{2}\right] z^{2}(1-z)\frac{2q_\perp^2}{\tilde{P}_{\perp }^2(\tilde{P}_\perp+zq_\perp)^2}\notag \\
&\times\int\frac{\text{d}^{2}v}{(2\pi)^{2}}\frac{\text{d}^{2}v^{\prime }}{(2\pi )^{2}}e^{-iq_{\perp}\cdot(v-v^{\prime })}S^{(2)}_{x_g}(v,v').
\end{align}
This result agrees with the factorized result (\ref{factphoton}) in the correlation limit $P_\perp\gg q_\perp$. To make more clear the relation between the distribution $xG^{(2)}$ in Eq. (\ref{g2}) and the result above notice that the factor $q_\perp^2$ can be brought inside the integral as derivatives of the exponential factor with respect to $v$ and $v'$. Using integration by parts and the derivation formula for Wilson lines it is easy to show that the cross section takes the form
\begin{align}
\frac{d\sigma^{pA\rightarrow \gamma q+X} }{d\mathcal{P.S.}}
=&\;\sum_{f}x_{p}q_{f}(x_{p})\alpha _{em}e_{f}^{2}\left[1+(1-z)^{2}\right] z^{2}(1-z)\frac{2}{\tilde{P}_{\perp }^2(\tilde{P}_\perp+zq_\perp)^2}\notag \\
&\times 16\pi^3\alpha_S\int\frac{\text{d}^{3}v}{(2\pi)^{3}}\frac{\text{d}^{3}v^{\prime }}{(2\pi )^{3}}e^{-iq_{\perp}\cdot(v-v^{\prime })}\left\langle\text{Tr}\left[F^{i-}(\vec{v})\mathcal{U}^{[-]\dagger}F^{i-}(\vec{v}')\mathcal{U}^{[+]}\right]\right\rangle.
\end{align}
Taking into account the same considerations about different normalizations of the averaging procedures as in the DIS case, it is easy to see that the two expressions for $xG^{(2)}$ agree in the small-$x$ region.

\section{Dijet production in $pA$ collisions}

Dijet production in $pA$ collisions receive contributions from
several channels such as $qg\to qg$, $gg\to q\bar{q}$ and $gg\to
gg$. For convenience, we define the following common variables as
in the last two sections,
\begin{equation}
z=\frac{|k_{1\perp}|e^{y_1}}{|k_{1\perp}|e^{y_1}+|k_{2\perp}|e^{y_2}}, \quad x_{p}=\frac{|k_{1\perp}|e^{y_1}+|k_{2\perp}|e^{y_2}}{\sqrt{s}}, \quad x_{g}=\frac{|k_{1\perp}|e^{-y_1}+|k_{2\perp}|e^{-y_2}}{\sqrt{s}}\ ,
\end{equation}
where $k_1$ and $k_2$ are momenta, and $y_1$ and $y_2$ are rapidities for the two
outgoing particles, $x_p$ is the momentum fraction of the projectile nucleon carried
by the incoming parton, $x_g$ is the momentum fraction of the target nucleus carried
by the gluon, respectively.
Taking into account that the quark distribution functions are dominant at large-$x$ and the gluon distribution functions are dominant at low-$x$, it comes as no surprise the fact that different channels are relevant in different kinematic regions. At RHIC energies, the low -$x$ region is only accessible in events where the two jets are produced in the forward rapidity region of the projectile. Under those conditions we have $x_p\sim0.1$ and $x_g\ll0.1$, and therefore quark initiated processes dominate ($qg\to qg$ channel).

The higher energies available at LHC will allow to explore more thoroughly the low-$x$ regime in the target nucleus as well as in the projectile(see e.g., in a recent study\cite{Deak:2010gk}). Under these circumstances, and in particular at central rapidities at the LHC, it is possible to have processes with both $x_p$ and $x_g$ small where the dominant channels are $gg\to q\bar{q}$ and $gg\to gg$.

Let us first take the partonic channel $qg\to qg$ as an example
and calculate the dijet production cross section. Then it is
straightforward to generalize the calculation to the other
partonic channels $gg\to q\bar{q}$ and $gg\to gg$.

\subsection{TMD-factorization approach}

\subsubsection{The $qg\to qg$ channel}

\begin{figure}[t]
\begin{center}
\includegraphics[width=13cm]{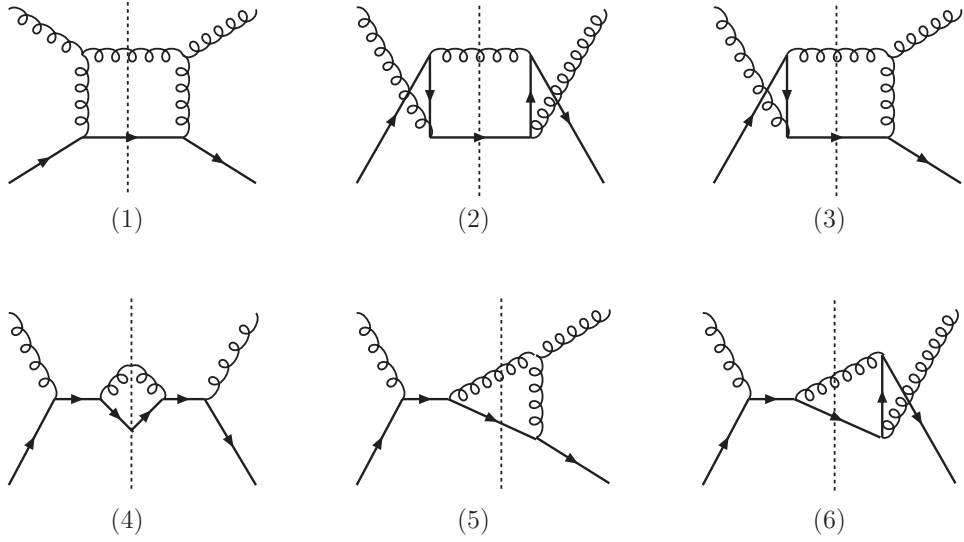}
\end{center}
\par
\vskip -0.4cm
\caption{Quark-gluon scattering diagrams. The mirror diagrams of
(3), (5) and (6) give identical contributions.}
\label{f10}
\end{figure}

%FY
The calculations follow the previous examples. However, there are
several different Feynman graphs contributing to the production of $qg$
in the final state, as shown in Fig.~\ref{f10}. In addition, they have different
color structures. Therefore, we need to compute the
hard factors and the associated initial/final state interaction
phases separately. In the end, we will sum their contributions together
to obtain the final result.

It is straightforward to obtain the hard cross section contributions
from each diagram in Fig.~\ref{f10} for the $qg\rightarrow qg$
process, and have been calculated in Ref.~\cite{qvy-short}.
We list these results in Table~\ref{qgtable} with the same notations,
where $h^{(i)}$ is the partonic hard factor and $C_u^{(i)}$ is the associated color
factor. In the calculations, in order to apply the eikonal approximation
when multiple gluon interactions are formulated, we have chosen the
physical polarizations for the outgoing gluon. However, the final result
for the differential cross section does not depend on this choice.
%FY

\begin{table}[t]
\caption{The color and hard factors for the $qg\rightarrow qg$ scattering
channels in Fig.~\ref{f10}, where $C_F=(N_c^2-1)/2N_c$. }
\begin{ruledtabular}
\begin{tabular}{|l|c|c|c|c|c|c|}
% \hline
& (1) & (2) & (3) & (4) & (5) & (6)  \\
\hline $h$ ~~~& $~~-\frac{4(\hat t^2-\hat s\hat u)^2}{\hat t^2\hat
s\hat u}~~$ &$~~-\frac{2(\hat u^2+\hat t^2)}{\hat s\hat u}~~$
&$~~\frac{2(\hat t^2-\hat s\hat u)(\hat u-\hat t)}{\hat s\hat
t\hat u}~~$ & $~~-\frac{2(\hat s^2+\hat t^2)}{\hat s\hat u}~~$ &
$~~-\frac{2(\hat t^2-\hat s\hat u)(\hat s-\hat t)}{\hat s\hat
t\hat u}~~$
&$~~\frac{2\hat t^2}{\hat s\hat u}~~$\\
\hline $C_u$  & $\frac{1}{2}$ &$\frac{C_F}{2N_c}$
       &$-\frac{1}{4}$ & $\frac{C_F}{2N_c}$& $\frac{1}{4}$ &
$-\frac{1}{4N_c^2}$
 \end{tabular}
\end{ruledtabular}
\label{qgtable}
\end{table}

As a consistency check, we can easily reproduce the known results for
the total hard cross section by summing all the graphs in Fig.~\ref{f10}
and explicitly taking $N_c=3$,
\begin{eqnarray}
\frac{\text{d}\hat{\sigma}}{\text{d}\hat{t}}\left( gq\rightarrow gq\right) &=&
\frac{g^{4}}{16\pi \hat{s}^{2}}\left\{\sum_{i=1,2,4} C_u^{(i)}h^{(i)}+
2\sum_{i=3,5,6} C_u^{(i)}h^{(i)}\right\}\nonumber\\
&=&\frac{g^{4}}{16\pi \hat{s}^{2}}\left( \frac{4}{9}\frac{\hat{s}^{2}+\hat{u}%
^{2}}{-\hat{s}\hat{u}}+\frac{\hat{s}^{2}+\hat{u}^{2}}{\hat{t}^{2}}\right) \ .
\end{eqnarray}
Since the graphs in Fig.~\ref{f10} have different color structure,
the gluon distributions associated with those graphs have
different gauge links according to Ref.~\cite{Bomhof:2006dp}.
Therefore, the corresponding gluon distributions in coordinate
space are found as follows:
\begin{eqnarray}
\Phi _{g}^{\left( 1\right) } &=&\left\langle \text{Tr}\left[ F\left( \xi
\right) \left\{ \frac{1}{2}\frac{\text{Tr}\left[ \mathcal{U}^{\left[ \square %
\right] }\right] }{N_{c}}\mathcal{U}^{\left[ +\right] \dagger }+\frac{1}{2}%
\mathcal{U}^{\left[ -\right] \dagger }\right\} F\left( 0\right) \mathcal{U}^{%
\left[ +\right] }\right] \right\rangle , \\
\Phi _{g}^{\left( 2\right) } &=&\left\langle \text{Tr}\left[ F\left( \xi
\right) \left\{ \frac{N_{c}^{2}}{N_{c}^{2}-1}\frac{\text{Tr}\left[ \mathcal{U%
}^{\left[ \square \right] }\right] }{N_{c}}\mathcal{U}^{\left[ +\right]
\dagger }-\frac{1}{N_{c}^{2}-1}\mathcal{U}^{\left[ -\right] \dagger
}\right\} F\left( 0\right) \mathcal{U}^{\left[ +\right] }\right]
\right\rangle , \\
\Phi _{g}^{\left( 3\right) } &=&\left\langle \text{Tr}\left[ F\left( \xi
\right) \frac{\text{Tr}\left[ \mathcal{U}^{\left[ \square \right] }\right] }{%
N_{c}}\mathcal{U}^{\left[ +\right] \dagger }F\left( 0\right) \mathcal{U}^{%
\left[ +\right] }\right] \right\rangle , \\
\Phi _{g}^{\left( 4\right) ,\left( 5\right) ,\left( 6\right) }
&=&\left\langle \text{Tr}\left[ F\left( \xi \right) \mathcal{U}^{\left[ -%
\right] \dagger }F\left( 0\right) \mathcal{U}^{\left[ +\right] }\right]
\right\rangle ,
\end{eqnarray}
where
$\mathcal{U}^{[\square]}=\mathcal{U}^{[+]}\mathcal{U}^{[-]^{\dagger}}=\mathcal{U}^{[-]^{\dagger}}\mathcal{U}^{[+]}$
emerges as a Wilson loop. Now we are ready to combine all the
channels together. As mentioned in the introduction, the distributions above will be factorizable in terms of convolutions of the two basic distributions from the previous sections. Anticipating this result, we consider only the leading
contribution in $N_c$. Noting that graph (6) in Fig~\ref{f10} does
not contribute in the large-$N_{c}$ limit, one can find
\begin{equation}
\frac{d\sigma _{\text{TMD}}^{qA\rightarrow qgX}}{d^{2}P_{\perp
}d^{2}q_{\perp }dy_{1}dy_{2}}=\sum_{f} x_{p}q(x_p)\frac{\alpha_s^2}{\hat{s}%
^{2}}\left[ \mathcal{F}_{qg}^{(1)}H_{qg\rightarrow qg}^{(1)}+\mathcal{F}%
_{qg}^{(2)}H_{qg\rightarrow qg}^{(2)}\right] ,\label{factqqg}
\end{equation}%
with
\begin{eqnarray}
\mathcal{F}_{qg}^{(1)} &=&xG^{(2)}\left(x, q_{\perp}\right)=2\int \frac{d\xi
^{-}d\xi _{\perp }}{(2\pi )^{3}P^{+}} e^{ixP^{+}\xi ^{-}-iq_{\perp }\cdot
\xi _{\perp }}\left\langle \text{Tr}\left[ F\left( \xi \right) \mathcal{U}^{%
\left[ -\right] \dagger }F\left( 0\right) \mathcal{U}^{\left[ +\right] }%
\right] \right\rangle , \label{Fqg1}\\
\mathcal{F}_{qg}^{(2)} &=&2\int \frac{d\xi ^{-}d\xi _{\perp }}{(2\pi
)^{3}P^{+}} e^{ixP^{+}\xi ^{-}-iq_{\perp }\cdot \xi _{\perp }}\left\langle
\text{Tr}\left[ F\left( \xi \right) \frac{\text{Tr}\left[ \mathcal{U}^{\left[
\square \right] }\right] }{N_{c}}\mathcal{U}^{\left[ +\right] \dagger
}F\left( 0\right) \mathcal{U}^{\left[ +\right] }\right] \right\rangle .\label{Fqg2}
\end{eqnarray}%
In the large-$N_{c}$ limit, it is straightforward to find that only graphs (1),
(2) and (3) in Fig.~\ref{f10} ($t$ and $u$ channels together with their cross diagrams)
contribute to $H_{qg\rightarrow qg}^{(2)}$ and only graphs (1), (4) and (5) ($t$ and $s$
channels together with their cross diagrams) contribute to $H_{qg\rightarrow qg}^{(1)}$.
By using $\frac{C_{F}}{2N_{c}}=\frac{1}{4}$ in the large-$N_{c}$ limit, one obtains
\begin{eqnarray}
H_{qg\rightarrow qg}^{(1)}&=&-\frac{\left( \hat{t}^{2}-\hat{s}\hat{u}\right)
^{2}}{\hat{s}\hat{u}\hat{t}^{2}}-\frac{1}{2}\frac{\hat{t}^{2}+\hat{s}^{2}}{%
\hat{s}\hat{u}}-\frac{\left( \hat{t}^{2}-\hat{s}\hat{u}\right) \left( \hat{s}%
-\hat{t}\right) }{\hat{s}\hat{u}\hat{t}}=-\frac{\hat{u}^{2}\left( \hat{s}%
^{2}+\hat{u}^{2}\right) }{2\hat{s}\hat{u}\hat{t}^{2}}, \\
H_{qg\rightarrow qg}^{(2)} &=&-\frac{\left( \hat{t}^{2}-\hat{s}\hat{u}%
\right) ^{2}}{\hat{s}\hat{u}\hat{t}^{2}}-\frac{1}{2}\frac{\hat{t}^{2}+\hat{u}%
^{2}}{\hat{s}\hat{u}}-\frac{\left( \hat{t}^{2}-\hat{s}\hat{u}\right) \left(
\hat{u}-\hat{t}\right) }{\hat{s}\hat{u}\hat{t}}=-\frac{\hat{s}^{2}\left(
\hat{s}^{2}+\hat{u}^{2}\right) }{2\hat{s}\hat{u}\hat{t}^{2}}\ .
\end{eqnarray}
We note that although the individual diagram's contribution to the above
two hard factors depends on the polarization we choose for the outgoing
gluon, the final results for the hard factors do not depend on this choice.
This means the combination of Feynman graphs according to the relevant
color structure is gauge invariant. Similar conclusion has also been obtained
for the spin related observables calculated in Refs.~\cite{Bomhof:2006dp,{qvy-short}}.

Since one has $\hat{s}=\frac{P_{\perp }^{2}}{z\left( 1-z\right) }$, $\hat{u}=-\frac{%
P_{\perp }^{2}}{z}$ and $\hat{t}=-\frac{P_{\perp }^{2}}{1-z}$ in the correlation limit, Eq. (\ref{factqqg}) leads to the following cross section for $qg$ dijet production in $pA$ collisions
\begin{eqnarray}
&&\frac{d\sigma _{\text{TMD}}^{pA\rightarrow qgX}}{d^{2}P_{\perp
}d^{2}q_{\perp }dy_{1}dy_{2}}  \notag \\
&=&\sum_{f}x_{p}q_{f}(x_{p})\frac{\alpha_s^2}{ 2P_{\perp }^{4}}%
\left[ 1+\left( 1-z\right) ^{2}\right] \left( 1-z\right) \left[ \left(
1-z\right) ^{2}xG^{(2)}\left(x, q_{\perp}\right)+\mathcal{F}_{qg}^{(2)}%
\right] ,\label{factqqg2}
\end{eqnarray}
where $x_{p}q_{f}(x_{p})$ is the integrated quark distribution for the proton projectile.

\subsubsection{The $gg\to q\bar{q}$ channel}

\begin{figure}[t]
\begin{center}
\includegraphics[width=13cm]{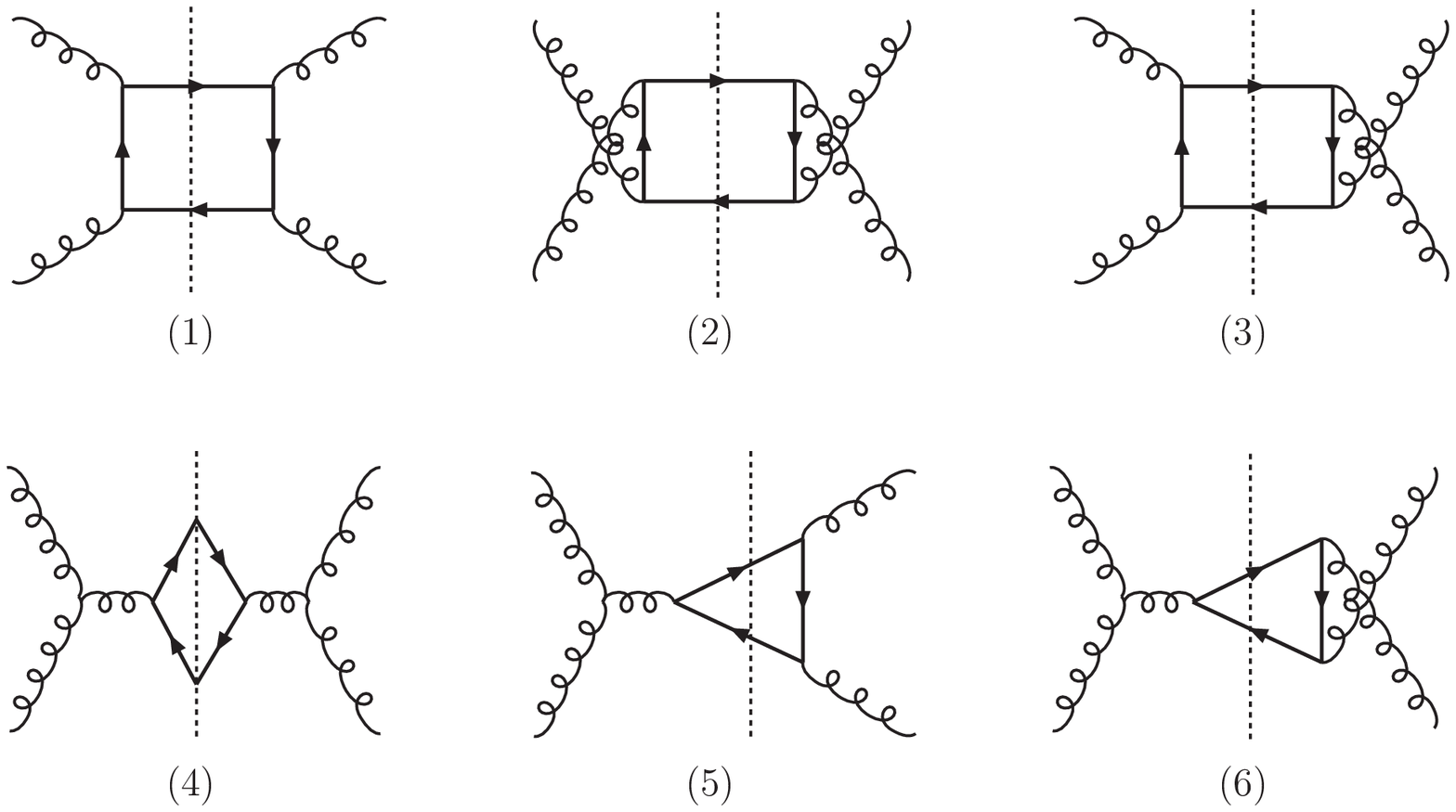}
\end{center}
\par
\vskip -0.4cm
\caption{$gg\to q\bar q$ scattering diagrams. The mirror diagrams of
(3), (5) and (6) give identical contributions.}
\label{fggqq}
\end{figure}

\begin{table}[t]
\caption{The color and hard factors for the $gg\rightarrow q\bar{q}$ scattering
channels in Fig.~\protect\ref{fggqq}. }%
\begin{ruledtabular}
\begin{tabular}{|l|c|c|c|c|c|c|}
% \hline
& (1) & (2) & (3) & (4) & (5) & (6)  \\
\hline $h$ ~~~&$\frac{2(3\hat t^2+\hat u^2)\hat u}{(\hat t+\hat u)^2\hat t}~~$&$\frac{2(\hat t^2+3\hat u^2)\hat t}{(\hat t+\hat u)^2\hat u}~~$
&$~~\frac{2(\hat t-\hat u)^2}{(\hat t+\hat u)^2}~~$ &
$~~\frac{4\hat t \hat u}{(\hat t+\hat u)^2}~~$ &
$~~-\frac{4\hat t \hat u}{(\hat t+\hat u)^2}~~$
&$~~\frac{4\hat t \hat u}{(\hat t+\hat u)^2}~~$\\
\hline $C_u$  & $\frac{1}{4N_c}$ &$\frac{1}{4N_c}$
       &$-\frac{1}{4N_c\left(N_c^2-1\right)}$ & $\frac{N_c}{2\left(N_c^2-1\right)}$& $\frac{N_c}{4\left(N_c^2-1\right)}$ & $-\frac{N_c}{4\left(N_c^2-1\right)}$
 \end{tabular}
\end{ruledtabular}
\label{tableggqq}
\end{table}

Following the same procedure illustrated in the $qg\to qg$ channel, we can calculate the dijet production cross section from the $gg\to q\bar{q}$ channel. First of all, we compute the color factors and hard factors for each graph in Fig.~\ref{fggqq} and list them in Table~\ref{tableggqq}. Then, we plug in the appropriate gluon distributions\footnote{We have simplified these gluon distributions by using large-$N_c$ limit and the fact that they are real in the CGC formalism. } as found in Ref.~\cite{Bomhof:2006dp}.
\begin{eqnarray}
\Phi _{g}^{\left( 1\right) ,\left( 2\right) } &=&\left\langle \text{Tr}\left[
F\left( \xi \right) \left\{ \frac{\text{Tr}\left[ \mathcal{U}^{\left[
\square \right] }\right] }{N_{c}}\mathcal{U}^{\left[ -\right] \dagger
}\right\} F\left( 0\right) \mathcal{U}^{\left[ +\right] }\right]
\right\rangle , \\
\Phi _{g}^{\left( 3\right) } &=&-N_{c}\left\langle \text{Tr}\left[ F\left(
\xi \right) \mathcal{U}^{\left[ \square \right] }\right] \text{Tr}\left[
F\left( 0\right) \mathcal{U}^{\left[ \square \right] \dagger }\right]
\right\rangle , \\
\Phi _{g}^{\left( 4\right) ,\left( 5\right) ,\left( 6\right) }
&=&\left\langle \text{Tr}\left[ F\left( \xi \right) \mathcal{U}^{\left[ -%
\right] \dagger }F\left( 0\right) \mathcal{U}^{\left[ +\right] }\right]
\frac{\text{Tr}\left[ \mathcal{U}^{\left[ \square \right] }\right] }{N_{c}}
\right\rangle \nonumber \\
&&-\frac{1}{N_{c}}\left\langle \text{Tr}\left[ F\left( \xi
\right) \mathcal{U}^{\left[ \square \right] }\right] \text{Tr}\left[ F\left(
0\right) \mathcal{U}^{\left[ \square \right] \dagger }\right] \right\rangle .
\end{eqnarray}
Combining all the channels in the large $N_c$ limit, we can find
\begin{equation}
\frac{d\sigma _{\text{TMD}}^{gA\rightarrow q\bar{q}X}}{d^{2}P_{\perp
}d^{2}q_{\perp }dy_{1}dy_{2}}=\sum_{f} x_{p}g(x_p)\frac{\alpha_s^2}{\hat{s}
^{2}}\left[ \mathcal{F}_{gg}^{(1)}H_{gg\rightarrow q\bar{q}}^{(1)}+\mathcal{F}
_{gg}^{(2)}H_{gg\rightarrow q\bar{q}}^{(2)}\right] , \label{ggtoqqbar}
\end{equation}%
with
\begin{eqnarray}
\mathcal{F}_{gg}^{(1)} &=&2\int \frac{d\xi
^{-}d\xi _{\perp }}{(2\pi )^{3}P^{+}} e^{ixP^{+}\xi ^{-}-iq_{\perp }\cdot
\xi _{\perp }}\left\langle \text{Tr}\left[ F\left( \xi \right)\frac{\text{Tr}\left[ \mathcal{U}^{\left[
\square \right] }\right] }{N_{c}} \mathcal{U}^{%
\left[ -\right] \dagger }F\left( 0\right) \mathcal{U}^{\left[ +\right] }%
\right] \right\rangle ,\label{fgg1} \\
\mathcal{F}_{gg}^{(2)} &=&2\int \frac{d\xi ^{-}d\xi _{\perp }}{(2\pi
)^{3}P^{+}} e^{ixP^{+}\xi ^{-}-iq_{\perp }\cdot \xi _{\perp }}\frac{1}{N_c}\left\langle
\textrm{Tr}\left[ F\left( \xi \right) \mathcal{U}^{\left[
\square\right]\dagger} \right]\textrm{Tr}\left[ F\left( 0\right) \mathcal{U}^{\left[ \square\right] }\right] \right\rangle \ ,\label{fgg2}
\end{eqnarray}
and
\begin{eqnarray}
H_{gg\rightarrow q\bar{q}}^{(1)} &=&\frac{1}{4N_c}\frac{2\left( \hat{t}^{2}+\hat{u}^{2}\right) ^{2}}{\hat{s}^{2}\hat{u}\hat{t}}\ ,\\
H_{gg\rightarrow q\bar{q}}^{(2)} &=&\frac{1}{4N_{c}}\frac{4\left( \hat{t}^{2}+\hat{u}^{2}\right) }{
\hat{s}^{2}},
\end{eqnarray}
where $x_{p}g(x_{p})$ is the integrated gluon distribution in the proton projectile.

\subsubsection{The $gg\to gg$ channel}

\begin{figure}[t]
\begin{center}
\includegraphics[width=13cm]{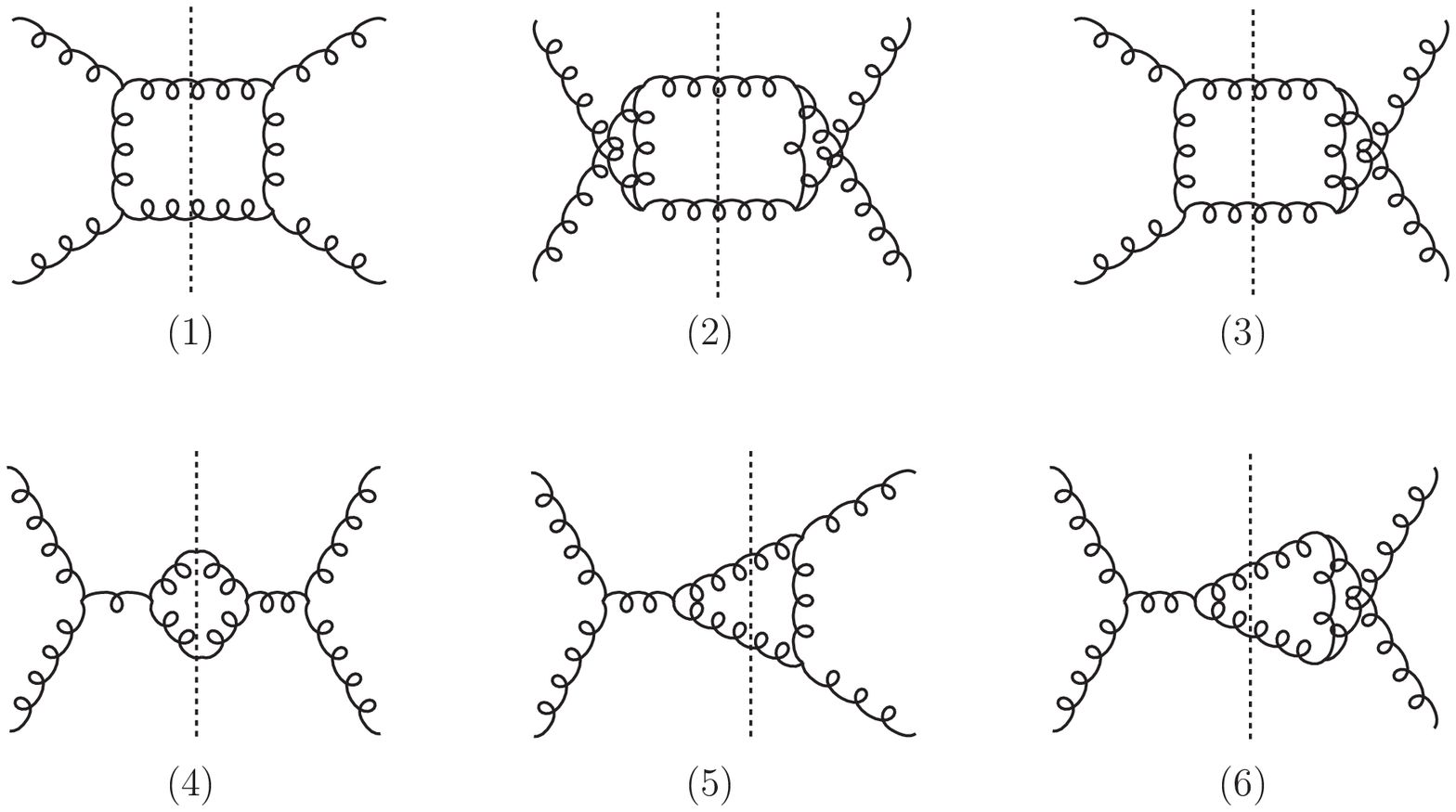}
\end{center}
\par
\vskip -0.4cm
\caption{$gg\to gg$ scattering diagrams. The mirror diagrams of
(3), (5) and (6) give identical contributions.}
\label{fgg}
\end{figure}

%\begin{table}[t]
%\caption{The color- and hard factors for the $gg\rightarrow gg$ scattering
%channels in Fig.~\protect\ref{fgg}, where $h^{(2)}_{gg\to gg}=\frac{2(2\hat{s}^6+6\hat s^5 \hat t+14\hat s^4\hat t^2+20\hat s^3 \hat t^3+21 \hat s^2 \hat t^4 +14 \hat s \hat t^5+4\hat %t^6)}{(\hat s +\hat t)^2 \hat s^2\hat t^2}$ and $h^{(3)}_{gg\to gg}=-\frac{(2\hat{s}^4+5\hat s^3 \hat t+10\hat s^2\hat t^2+10 \hat s \hat t^3+4\hat t^4)(\hat s +2 \hat t)}{(\hat s +\hat t)^2 %\hat s^2 \hat t}$. }%
%\begin{ruledtabular}
%\begin{tabular}{|l|c|c|c|c|c|c|}
% \hline
%& (1) & (2) & (3) & (4) & (5) & (6)  \\
%\hline $h$ & $\frac{2(\hat{s}^4+4\hat s^3 \hat t+11\hat s^2\hat t^2+10 \hat s \hat t^3+4\hat t^4)}{(\hat s +\hat t)^2 \hat s^2}$ &$h^{(2)}_{gg\to gg}$
%&$h^{(3)}_{gg\to gg}$ & $\frac{2(\hat{s}^4+\hat s^3 \hat t+5\hat s^2\hat t^2+6\hat s \hat t^3+2\hat t^4)}{(\hat s +\hat t)^2 \hat s^2}$ &
%$\frac{2\hat s^5+ \hat s^4 \hat t -\hat s^3 \hat t^2- 10 \hat s^2 \hat t^3- 12 \hat s \hat t^4 -4 \hat t^5 }{(\hat s+\hat t)^2\hat s^2 \hat t}$
%&$\frac{(\hat s^3+10\hat s^2\hat t+12\hat s \hat t^2+4\hat t^3)t}{(\hat s +\hat t)^2
%\hat s^2}$\\
%\hline $C_u$  & $\frac{N_c^2}{N_c^2-1}$ &$\frac{N_c^2}{N_c^2-1}$
%       &$\frac{N_c^2}{2(N_c^2-1)}$ & $\frac{N_c^2}{N_c^2-1}$& $\frac{N_c^2}{2(N_c^2-1)}$ &
%$-\frac{N_c^2}{2(N_c^2-1)}$
 %\end{tabular}
%\end{ruledtabular}
%\end{table}

\begin{table}[t]
\caption{The color and hard factors for the $gg\rightarrow gg$ scattering
channels in Fig.~\ref{fgg}.}%
\begin{ruledtabular}
\begin{tabular}{|c|c|c|}
% \hline
 & h &$C_u$ \\
\hline (1) & $\frac{2(\hat{s}^4+4\hat s^3 \hat t+11\hat s^2\hat t^2+10 \hat s \hat t^3+4\hat t^4)}{(\hat s +\hat t)^2 \hat s^2}$ & $\frac{N_c^2}{N_c^2-1}$ \\
\hline (2) & $ \frac{2(2\hat{s}^6+6\hat s^5 \hat t+14\hat s^4\hat t^2+20\hat s^3 \hat t^3+21 \hat s^2 \hat t^4 +14 \hat s \hat t^5+4\hat t^6)}{(\hat s +\hat t)^2 \hat s^2\hat t^2}$  & $\frac{N_c^2}{N_c^2-1}$ \\
\hline (3) & $-\frac{(2\hat{s}^4+5\hat s^3 \hat t+10\hat s^2\hat t^2+10 \hat s \hat t^3+4\hat t^4)(\hat s +2 \hat t)}{(\hat s +\hat t)^2 \hat s^2 \hat t}$ & $\frac{N_c^2}{2(N_c^2-1)}$\\
\hline (4) & $\frac{2(\hat{s}^4+\hat s^3 \hat t+5\hat s^2\hat t^2+6\hat s \hat t^3+2\hat t^4)}{(\hat s +\hat t)^2 \hat s^2}$  & $\frac{N_c^2}{N_c^2-1}$\\
\hline (5) & $\frac{2\hat s^5+ \hat s^4 \hat t -\hat s^3 \hat t^2- 10 \hat s^2 \hat t^3- 12 \hat s \hat t^4 -4 \hat t^5 }{(\hat s+\hat t)^2\hat s^2 \hat t}$ & $\frac{N_c^2}{2(N_c^2-1)}$ \\
\hline (6) & $\frac{(\hat s^3+10\hat s^2\hat t+12\hat s \hat t^2+4\hat t^3)t}{(\hat s +\hat t)^2
\hat s^2}$ & $-\frac{N_c^2}{2(N_c^2-1)}$
 \end{tabular}
\end{ruledtabular}
\label{tablegg}
\end{table}

Similarly, the color factors and hard factors for all the graphs
plotted in Fig.~\ref{fgg} have been calculated and listed in
Table~\ref{tablegg}. Combining these factors with the
corresponding gluon distributions, taking into account the
appropriate gauge links~\cite{Bomhof:2006dp}, we arrive at
\begin{eqnarray}
\Phi _{g}^{\left( 1\right) ,\left( 2\right) } &=&\frac{1}{2}\left\langle
\text{Tr}\left[ F\left( \xi \right) \mathcal{U}^{\left[ +\right] \dagger
}F\left( 0\right) \mathcal{U}^{\left[ +\right] }\right] \frac{\text{Tr}\left[
\mathcal{U}^{\left[ \square \right] }\right] }{N_{c}}\frac{\text{Tr}\left[
\mathcal{U}^{\left[ \square \right] }\right] }{N_{c}}\right\rangle\nonumber \\
&&+\left\langle \text{Tr}\left[ F\left( \xi \right) \mathcal{U}^{\left[ -%
\right] \dagger }F\left( 0\right) \mathcal{U}^{\left[ +\right] }\right]
\frac{\text{Tr}\left[ \mathcal{U}^{\left[ \square \right] }\right] }{N_{c}}%
\right\rangle , \\
\Phi _{g}^{\left( 3\right) } &=&\left\langle \text{Tr}\left[ F\left( \xi
\right) \mathcal{U}^{\left[ +\right] \dagger }F\left( 0\right) \mathcal{U}^{%
\left[ +\right] }\right] \frac{\text{Tr}\left[ \mathcal{U}^{\left[ \square %
\right] }\right] }{N_{c}}\frac{\text{Tr}\left[ \mathcal{U}^{\left[ \square %
\right] }\right] }{N_{c}}\right\rangle \nonumber \\
&&+\frac{1}{N_{c}}\left\langle \text{Tr}\left[ F\left( \xi \right) \mathcal{U%
}^{\left[ \square \right] }\right] \text{Tr}\left[ F\left( 0\right) \mathcal{%
U}^{\left[ \square \right] \dagger }\right] \right\rangle , \\
\Phi _{g}^{\left( 4\right) ,\left( 5\right) ,\left( 6\right) }
&=&\left\langle \text{Tr}\left[ F\left( \xi \right) \mathcal{U}^{\left[ -%
\right] \dagger }F\left( 0\right) \mathcal{U}^{\left[ +\right] }\right]
\frac{\text{Tr}\left[ \mathcal{U}^{\left[ \square \right] }\right] }{N_{c}}%
\right\rangle  \nonumber \\
&&-\frac{1}{N_{c}}\left\langle \text{Tr}\left[ F\left( \xi \right) \mathcal{U%
}^{\left[ \square \right] }\right] \text{Tr}\left[ F\left( 0\right) \mathcal{%
U}^{\left[ \square \right] \dagger }\right] \right\rangle .
\end{eqnarray}
Summing over all the channels in the large-$N_c$ limit, we can obtain
\begin{equation}
\frac{d\sigma _{\text{TMD}}^{gA\rightarrow gg X}}{d^{2}P_{\perp
}d^{2}q_{\perp }dy_{1}dy_{2}}=\sum_{f} x_{p}g(x_p)\frac{\alpha_s^2}{\hat{s}
^{2}}\left[ \mathcal{F}_{gg}^{(1)}H_{gg\rightarrow gg}^{(1)}+\mathcal{F}
_{gg}^{(2)}H_{gg\rightarrow gg}^{(2)}+\mathcal{F}_{gg}^{(3)}H_{gg\rightarrow gg}^{(3)}\right] ,\label{factggg}
\end{equation}%
where ${\cal F}_{gg}^{(1,2)}$ have been defined in Eqs.~(\ref{fgg1},\ref{fgg2}) and
${\cal F}_{gg}^{(3)}$ is defined as
\begin{eqnarray}
\mathcal{F}_{gg}^{(3)} &=&2\int \frac{d\xi
^{-}d\xi _{\perp }}{(2\pi )^{3}P^{+}} e^{ixP^{+}\xi ^{-}-iq_{\perp }\cdot
\xi _{\perp }}\left\langle \text{Tr}\left[ F\left( \xi \right)\mathcal{U}^{%
\left[ -\right] \dagger }F\left( 0\right) \mathcal{U}^{\left[ +\right] }%
\right] \frac{\text{Tr}\left[ \mathcal{U}^{\left[
\square \right] }\right] }{N_{c}} \frac{\text{Tr}\left[ \mathcal{U}^{\left[
\square \right] }\right] }{N_{c}}\right\rangle \ .\label{fgg3}
\end{eqnarray}
The hard factors are found as
\begin{eqnarray}
H_{gg\rightarrow gg}^{(1)} &=&\frac{2\left( \hat{t}^{2}+\hat{u}^{2}\right)
\left( \hat{s}^{2}-\hat{t}\hat{u}\right) ^{2}}{\hat{u}^{2}\hat{t}^{2}\hat{s}%
^{2}} , ~~ H_{gg\rightarrow gg}^{(2)} =\frac{4\left( \hat{s}^{2}-\hat{t}\hat{%
u}\right) ^{2}}{\hat{u}\hat{t}\hat{s}^{2}}  \ ,\notag \\
H_{gg\rightarrow gg}^{(3)} &=&\frac{2\left( \hat{s}^{2}-\hat{t}\hat{u}%
\right) ^{2}}{\hat{u}^{2}\hat{t}^{2}}.
\end{eqnarray}

Using the mean field approximation~\cite{Xiao:2010sp}, we can
simplify the gluon distributions and find the total dijet
production cross section which includes the $qg\to qg$, $gg\to
q\bar{q}$ and $gg \to gg$ channels as follows
\begin{eqnarray}
\frac{d\sigma^{(pA\rightarrow \mathrm{Dijet}+X)}}{d\mathcal{P.S.}}
&=&\sum_{q}x_{1}q(x_{1})\frac{\alpha_s^2}{\hat{s}^{2}}\left[ \mathcal{F}%
_{qg}^{(1)}H_{qg\rightarrow qg}^{(1)}+\mathcal{F}_{qg}^{(2)}H_{qg\rightarrow
qg}^{(2)}\right]  \notag \\
&&~~+x_{1}g(x_{1})\frac{\alpha_s^2}{\hat{s}^{2}}\left[ \mathcal{F}%
_{gg}^{(1)}\left( H_{gg\rightarrow
q\bar{q}}^{(1)}+\frac{1}{2}H_{gg\rightarrow
gg}^{(1)}\right) \right.  \notag \\
&&~~\left. +\mathcal{F}_{gg}^{(2)}\left( H_{gg\rightarrow q\bar{q}%
}^{(2)}+\frac{1}{2}H_{gg\rightarrow gg}^{(2)}\right) +\frac{1}{2}\mathcal{F}_{gg}^{(3)}H_{gg\rightarrow gg}^{(3)}\right] \ ,  \label{dijet}
\end{eqnarray}
where again $q(x_1)$ and $g(x_1)$ are integrated quark and gluon
distributions from the projectile nucleon. We have included a
statistical factor of $\frac{1}{2}$ in Eq.~(\ref{dijet}) for the
$gg\to gg$ channel due to the identical final state. The various
gluon distributions of nucleus $A$ are defined as
\begin{eqnarray}
\mathcal{F}_{qg}^{(1)} &=&xG^{(2)}(x,q_{\perp }),~~\mathcal{F}%
_{qg}^{(2)}=\int xG^{(1)}(q_1)\otimes F(q_2)\ ,  \notag \\
\mathcal{F}_{gg}^{(1)} &=&\int xG^{(2)}(q_1)\otimes F(q_2),~~\mathcal{F}%
_{gg}^{(2)}=-\int \frac{q_{1\perp}\cdot q_{2\perp}}{q_{1\perp}^2}
xG^{(2)}(q_1)\otimes F(q_2)\ ,  \notag \\
\mathcal{F}_{gg}^{(3)} &=&\int xG^{(1)}(q_{1})\otimes F(q_2)\otimes F(q_3)\ ,
\end{eqnarray}%
where $\otimes $ represents the convolution in momentum space: $\int
\otimes=\int d^2q_{1}d^2q_2\delta^{(2)}(q_\perp-q_1-q_2)$. These expressions follow directly from Eqs. (\ref{Fqg1}), (\ref{Fqg2}), (\ref{fgg1}), (\ref{fgg2}), (\ref{fgg3}) and the assumption that in the large-$N_c$ limit the expectation values involved in these equations can be factored as products of expectation values of the traces within. Clearly, this
process depends on both UGDs in a complicated way, and the naive TMD%
-factorization does not hold.

\subsection{CGC Calculations}

\subsubsection{$q\to qg$}

\begin{figure}[tbp]
\begin{center}
\includegraphics[width=9cm]{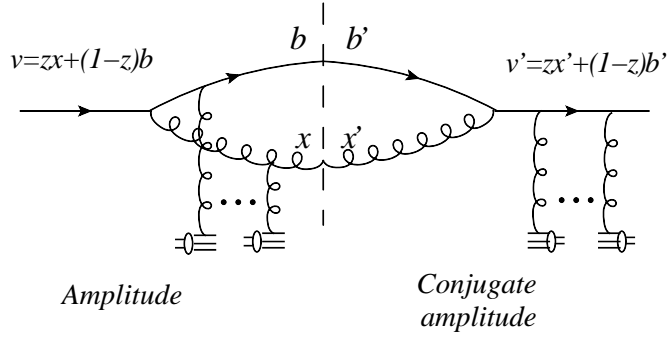} \\[0pt]
\end{center}
\caption[*]{Interactions before and after the splitting have to be taken into account for both amplitude and conjugate amplitude. After the splitting the nucleus interacts coherently with the quark-gluon system. Here is a typical diagram representing the second interaction term in Eq. (\ref{partqqg}).}
\label{qqgcgc}
\end{figure}

This process is studied in detail in Refs.~\cite{{JalilianMarian:2004da},Marquet:2007vb}, and
in particular Ref.~\cite{Marquet:2007vb} is close to the approach we have followed in this paper,
where an explicit formula analogous to the ones cited here for DIS and photon emission is given. We take as starting point Eq.~(24) in Ref.~\cite{Marquet:2007vb} which in our notation takes the form (see Fig. \ref{qqgcgc})
\begin{eqnarray}
\frac{d\sigma ^{qA\rightarrow qg X}}{d^3k_1d^3k_2}&=&\alpha _{S}C_F\delta(p^+-k_1^+-k_2^+) \int \frac{\text{d}^{2}x}{(2\pi)^{2}}\frac{\text{d}^{2}x^{\prime }}{(2\pi )^{2}}\frac{\text{d}^{2}b}{(2\pi)^{2}}\frac{\text{d}^{2}b^{\prime }}{(2\pi )^{2}}  \notag \\
&&\times e^{-ik_{1\perp }\cdot(x-x^{\prime })}e^{-ik_{2\perp }\cdot (b-b^{\prime })} \sum_{\lambda\alpha\beta} \psi^{\lambda\ast}_{\alpha\beta}(x'-b')\psi^\lambda_{\alpha\beta}(x-b)  \notag \\
&&\times \left[S^{(6)}_{x_g}(b,x,b',x')-S^{(3)}_{x_g}(b,x,zx'+(1-z)b')\right.\notag \\
&&\quad\left.-S^{(3)}_{x_g}(zx+(1-z)b,x',b')+S^{(2)}_{x_g}(zx+(1-z)b,zx'+(1-z)b')\right].\label{partqqg}
\end{eqnarray}
where
\begin{align}
S^{(6)}_{x_g}(b,x,b',x')&=\frac{1}{C_FN_c}\left\langle\text{Tr}\left(U(b)U^\dagger(b')T^dT^c\right)\left[W(x)W^\dagger(x')\right]^{cd}\right\rangle_{x_g},\\
S^{(3)}_{x_g}(b,x,v')&=\frac{1}{C_FN_c}\left\langle\text{Tr}\left(U(b)T^dU^\dagger(v')T^c\right)W^{cd}(x)\right\rangle_{x_g},
\end{align}
and $W(x)$ is a Wilson line in the adjoint representation. In the correlators above, Wilson lines in the fundamental representation appear when considering the multiple interaction of a quark with the nucleus and Wilson lines in the adjoint representation appear when considering multiple interactions of a gluon with the nucleus. Clearly, the $S^{(6)}_{x_g}$ term represents the case where interactions occur after the splitting both in the amplitude and in the conjugate amplitude, the $S^{(3)}_{x_g}$ terms represent the interference terms, and the $S^{(2)}_{x_g}$ term represent interactions before the splitting only.

This formula for the cross section has the same structure as Eqs. (\ref{xsdis}) and (\ref{xsphoton}). The splitting wave function is the same as in the photon emission case (Eq. (\ref{wvfunction})). The only difference in the emission vertex is a color matrix which is included as part of the multiple scattering factor (therefore confining the color algebra to just the multiple scattering factor). Using Fierz identities, the terms appearing in the multiple scattering factor above can be written in terms of fundamental Wilson lines only as
\begin{align}
S^{(6)}_{x_g}(b,x,b',x')&=\frac{1}{2C_FN_c}\left\langle\text{Tr}\left(U(b)U^\dagger(b')U(x')U^\dagger(x)\right)\text{Tr}U(x)U^\dagger(x')-\frac{1}{N_c}\text{Tr}U(b)U^\dagger(b')\right\rangle_{x_g},\\
S^{(3)}_{x_g}(b,x,v')&=\frac{1}{2C_FN_c}\left\langle\text{Tr}U(b)U^\dagger(x)\text{Tr}U(x)U^\dagger(v')-\frac{1}{N_c}\text{Tr}U(b)U^\dagger(v')\right\rangle_{x_g}, \label{s3}
\end{align}

Some of the correlators appearing in the expressions above are
familiar or have been calculated in the literature before. The
4-point function in $S^{(3)}_{x_g}$ is different from the one
appearing in the DIS case but it has been studied and calculated
in a model with Gaussian distribution of sources in
\cite{Dominguez:2008aa}. The 6-point function appearing in
$S^{(6)}_{x_g}$ presents a more difficult challenge even with only
four independent coordinates. In order to deal with this
difficulty, it is convenient to address the problem in the
large-$N_c$ limit where correlators of products of traces are
evaluated as product of correlators of traces. Specifically, for
the correlators above we get
\begin{align}
S^{(6)}_{x_g}(b,x,b',x')&\simeq\frac{1}{N_c^2}\left\langle\text{Tr}U(b)U^\dagger(b')U(x')U^\dagger(x)\right\rangle_{x_g}\left\langle\text{Tr}U(x)U^\dagger(x')\right\rangle_{x_g},\\
&=S^{(4)}_{x_g}(b,x,b',x')S^{(2)}_{x_g}(x,x'),\\
S^{(3)}_{x_g}(b,x,v')&\simeq\frac{1}{N_c^2}\left\langle\text{Tr}U(b)U^\dagger(x)\right\rangle_{x_g}\left\langle\text{Tr}U(x)U^\dagger(v')\right\rangle_{x_g},\\
&=S^{(2)}_{x_g}(b,x)S^{(2)}_{x_g}(x,v').
\end{align}
Note that in the large-$N_c$ limit, the 6-point function is related to the 4-point function that appeared in the DIS dijet case. In Appendix~\ref{4point}, we point out that the result of \cite{Marquet:2007vb} for the
6-point function misses the inelastic part of $S^{(4)}_{x_g}$.

To enforce the correlation limit we follow the procedure used in the DIS case. From the structure of the terms in the multiple scattering factor, we can see that the same kind of cancellations will occur and the final result will be the sum of the lowest order non-vanishing terms from the expansion of $S^{(6)}_{x_g}$. Moreover, since there is no linear term in the expansion of $S^{(4)}_{x_g}$, the lowest non-vanishing terms come separately from the $S^{(4)}_{x_g}$ factor and the $S^{(2)}_{x_g}$ in the same fashion as in the previous calculations for DIS and photon emission. With the previous considerations in mind, it is easy to see that the final result takes the form
\begin{equation}
\begin{split}
\frac{d\sigma^{pA\to qgX}}{d^2q_\perp d^2P_\perp dy_1dy_2}=&\;\sum_fx_pq_f(x_p)16\pi^3\frac{\alpha_S^2}{P_\perp^4}(1-z)\left[1+(1-z)^2\right]\\
&\times\int\frac{d^3v}{(2\pi)^3}\frac{d^3v'}{(2\pi)^3}e^{-iq_\perp\cdot(v-v')}\left[(1-z)^2\left\langle\text{Tr}\left[F^{i-}(\vec{v})\mathcal{U}^{[-]\dagger}F^{i-}(\vec{v}')\mathcal{U}^{[+]}\right]\right\rangle_{x_g}\right. \\
&\qquad\left.+\frac{1}{N_c}\left\langle\text{Tr}U(v)U^\dagger(v')\right\rangle_{x_g}\left\langle\text{Tr}\left[F^{i-}(\vec{v})\mathcal{U}^{[+]\dagger}F^{i-}(\vec{v}')\mathcal{U}^{[+]}\right]\right\rangle_{x_g}\right].\label{cswilson}
\end{split}
\end{equation}

Taking into account the difference between the normalizations, it is straightforward to see that the result above agrees with the factorized formula (\ref{factqqg2}).

\begin{figure}[tbp]
\begin{center}
\includegraphics[width=9cm]{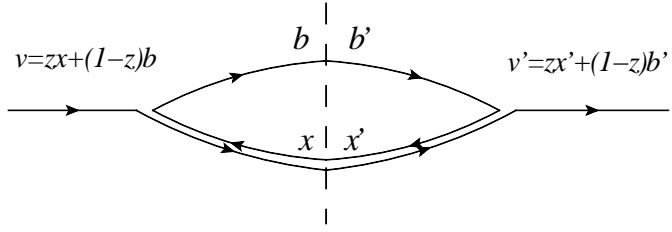} \\[0pt]
\end{center}
\caption[*]{Graphical representation of the splitting $q\to qg$ in the large-$N_c$ limit, in the amplitude and the conjugate amplitude.}
\label{lncqqg}
\end{figure}

In order to bring some insight to the relation between the
processes considered so far, and how the different distributions
come in for this particular channel, it is useful to consider the
graphical representation of the large-$N_c$ limit used to
factorize the correlators of Wilson lines. In the large-$N_c$
limit, a gluon line can be effectively considered as a
quark-antiquark pair. Forgetting about the multiple interactions
for the moment, and focusing primarily in the color flow of the
process, we see that in the large-$N_c$ limit the process takes
the form depicted in Fig. \ref{lncqqg}. The system splits into two
separate pieces, a quark line in the lower part of the diagram
which resembles the photon emission process, and a loop in the
upper part of the diagram which has the same color structure as
the DIS dijet case. Interactions involving both parts of the
process are $N_c$-suppressed, so it comes as no surprise that the
final result can be written as two separate pieces each involving
the respective distributions found for DIS and photon emission.

The fact that one of the terms in the final result involves only
one of the distributions while the other involves a convolution of
two factors can also be understood in a simple way from the
previous considerations. The enforcement of the correlation limit
is schematically the same as singling out one hard scattering in
the process and then taking $u=u'=0$ for the rest of the
interactions. When the hard scattering occurs on the lower part of
the diagram in Fig. \ref{lncqqg}, the quark-antiquark pair in the
upper part does not interact by color transparency
($S^{(4)}_{x_g}(b,b;b',b')=1$) and therefore there is no trace of
it in the first term of the factorized expression. When the hard
scattering occurs in the upper part of the diagram in Fig.
\ref{lncqqg}, the quark in the lower part still interacts with the
nucleus (and exchanges transverse momentum) and therefore has to
be included in the form of a dipole cross section.

\subsubsection{$g\to q\bar{q}$}

Following the same strategy from previous sections, we start with
the partonic level formula for the cross section built from the
splitting wave function and the multiple scattering factor. In
this particular case it takes the following form,
\begin{eqnarray}
\frac{d\sigma ^{gA\rightarrow q\bar{q}X}}{d^3k_1d^3k_2}&=&\alpha _S\delta(p^+-k_1^+-k_2^+)\frac{1}{2} \int \frac{\text{d}^{2}x_1}{(2\pi)^{2}}\frac{\text{d}^{2}x_1^{\prime }}{(2\pi )^{2}}\frac{\text{d}^{2}x_2}{(2\pi)^{2}}\frac{\text{d}^{2}x_2^{\prime }}{(2\pi )^{2}} \notag \\
&&\times e^{-ik_{1\perp }\cdot(x_1-x_1^{\prime })} e^{-ik_{2\perp }\cdot (x_2-x_2^{\prime })} \sum_{\lambda\alpha\beta} \psi_{\alpha\beta}^{T\lambda}(x_1-x_2)\psi_{\alpha\beta}^{T\lambda*}(x_1^{\prime }-x_2^{\prime })  \notag \\
&&\times \left[C_{x_g}(x_1,x_2,x'_1,x'_2)+S^A_{x_g}(zx_1+(1-z)x_2,zx'_1+(1-z)x'_2)\right.\notag\\
&&\quad\left.-S^{(3)}_{x_g}(x_1,zx'_1+(1-z)x'_2,x_2)-S^{(3)}_{x_g}(x'_2,zx_1+(1-z)x_2,x'_1)\right] \ ,\label{xsgqqbar}
\end{eqnarray}
where $S^{(3)}_{x_g}$ is given by (\ref{s3}) and
\begin{align}
C_{x_g}(x_1,x_2,x'_1,x'_2)&=\frac{1}{C_FN_c}\left\langle\text{Tr}\left(U^\dagger(x_2)T^cU(x_1)U^\dagger(x'_1)T^cU(x'_2)\right)\right\rangle_{x_g},\\
S^A_{x_g}(v,v')&=\frac{1}{N_c^2-1}\left\langle\text{Tr}W(v)W^\dagger(v')\right\rangle,
\end{align}
and the splitting wave function is the same as in the DIS case with $Q^2=0$. Notice this cross section is down by a factor of $N_c$ as compared to the $q\to qg$ case. This is due to the averaging over the incoming particle which amounts for a factor of $\frac{1}{N_c^2-1}$ for gluons instead of the factor of $\frac{1}{N_c}$ for quarks.

All the correlators above have been previously studied in the literature and explicit expressions for a Gaussian distribution of charges have been found. The only new ingredient that has not been considered in previous sections is $C_{x_g}$ which was thoroughly studied in \cite{Blaizot:2004wv}. Following the procedure from the previous section, let us express the correlators defined above in terms of fundamental Wilson lines only by means of Fierz identities.
\begin{align}
C_{x_g}(x_1,x_2,x'_1,x'_2)&=\frac{1}{2C_FN_c}\left\langle\text{Tr}U(x_1)U^\dagger(x'_1)\text{Tr}U(x'_2)U^\dagger(x_2)\right.\nonumber\\
&\left.~~~-\frac{1}{N_c}\text{Tr}U(x_1)U^\dagger(x'_1)U(x'_2)U^\dagger(x_2)\right\rangle_{x_g},\\
S^A_{x_g}(v,v')&=\frac{1}{N_c^2-1}\left\langle\text{Tr}U(v)U^\dagger(v')\text{Tr}U(v')U^\dagger(v)-1\right\rangle_{x_g}.
\end{align}
We take the large-$N_c$ limit in order to be able to compare with the results from the previous section and relate the cross section to the gluon distributions defined before. Under this approximation, the correlators above can be expressed entirely in terms of 2-point functions.
\begin{align}
C_{x_g}(x_1,x_2,x'_1,x'_2)&\simeq S^{(2)}_{x_g}(x_1,x'_1)S^{(2)}_{x_g}(x'_2,x_2),\\
S^A_{x_g}(v,v')&\simeq S^{(2)}_{x_g}(v,v')S^{(2)}_{x_g}(v',v).
\end{align}
This way of factorizing the correlators and the fact that the 4-point function is absent suggests that this process is related to the distribution given by the Fourier transform of the dipole cross section only. With this consideration in mind, we Fourier transform all the $S^{(2)}_{x_g}$ factors and perform the usual change of variables $u=x_1-x_2$ and $v=zx_1+(1-z)x_2$ (and similarly for the primed coordinates) and obtain
\begin{eqnarray}
\frac{d\sigma ^{gA\rightarrow q\bar{q} X}}{d^3k_1d^3k_2}&=&\alpha _{S}\delta(p^+-k_1^+-k_2^+) \frac{1}{2}\int \frac{\text{d}^{2}u}{(2\pi)^{2}}\frac{\text{d}^{2}u^{\prime }}{(2\pi )^{2}}\frac{\text{d}^{2}v}{(2\pi)^{2}}\frac{\text{d}^{2}v^{\prime }}{(2\pi )^{2}}\text{d}^2q_1\text{d}^2q_2 F_{x_g}(q_1)F_{x_g}(q_2)  \notag \\
&&\times e^{-i(q_{\perp}-q_1-q_2)\cdot(v-v^{\prime })}e^{-i\tilde{P}_\perp\cdot(u-u')} \sum_{\lambda\alpha\beta} \psi^{\lambda\ast}_{\alpha\beta}(u')\psi^\lambda_{\alpha\beta}(u)  \notag \\
&&\times \left[e^{i((1-z)q_2-zq_1)\cdot(u-u')}-e^{i((1-z)q_2-zq_1)\cdot u}-e^{-i((1-z)q_2-zq_1)\cdot u'}+1\right].
\end{eqnarray}

As in the photon emission case, the $u$ and $u'$ integrations reduce to calculate the Fourier transform of the splitting wave function with different momentum variables for each of the terms. The $v$ and $v'$ integrations give a $\delta$-function relating the momentum variables of the two distributions and a factor of the total transverse area. As in previous cases, we use collinear factorization for the incoming parton from the proton projectile and obtain
\begin{align}
\frac{d\sigma^{pA\rightarrow q\bar{q}X} }{d\mathcal{P.S.}}=&\;x_{p}g_{f}(x_{p})\alpha_S\left[z^2+(1-z)^{2}\right] z(1-z)\frac{S_\perp}{(2\pi)^2}\notag \\
&\times\int\text{d}^2q_1\text{d}^2q_2\delta^{(2)}(q_\perp-q_1-q_2)F_{x_g}(q_1)F_{x_g}(q_2)\frac{(zq_1-(1-z)q_2)^2}{\tilde{P}_{\perp }^2(\tilde{P}_\perp+zq_1-(1-z)q_2)^2}.
\end{align}
In the correlation limit, the denominator of the last fraction above becomes just $P_\perp^4$. From this expression it is clear that the distributions involved will be written as a convolution of two factors involving the Fourier transform of the dipole cross section. To notice how this equation above agrees with the factorized form in (\ref{ggtoqqbar}), expand the numerator and write the momentum factors as derivatives with respect to transverse coordinates of the dipole cross sections inside the definition of $F_{x_g}$ as was explained for the case of photon emission. There is a subtlety concerning the relative signs of the different terms when this identification is made. In order to find a complete agreement between the formula above and the factorized formula from the TMD formalism, it is necessary to write the two $F_{x_g}$ factors as Fourier transforms of Wilson loops in opposite directions (one of them in terms of $\mathcal{U}^{[\square]}$ and the other in terms of $\mathcal{U}^{[\square]\dagger}$). Because of this, $q_1$ and $q_2$ enter with opposite signs when expressed as derivatives of the Wilson loops. This sign is not visible in the terms with $q_1^2$ or $q_2^2$ but it changes the sign of the cross term, giving complete agreement with the factorized expression.

\begin{figure}[tbp]
\begin{center}
\includegraphics[width=9cm]{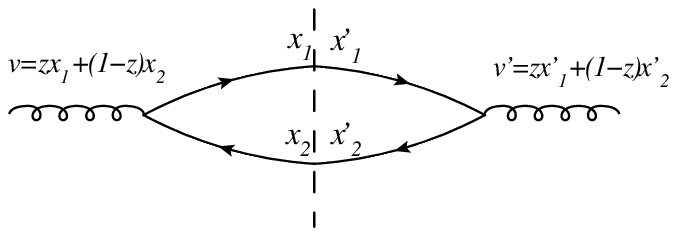} \\[0pt]
\includegraphics[width=9cm]{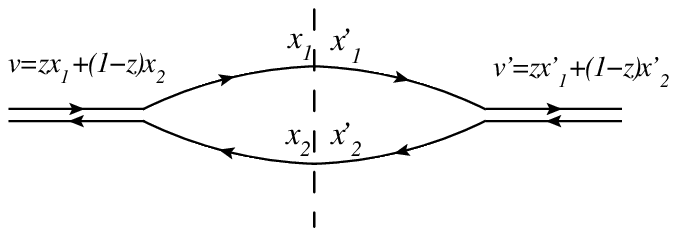} \\[0pt]
\end{center}
\caption[*]{Above: graphical representation of the splitting in the amplitude and conjugate amplitude. Below: splitting in the large-$N_c$ limit.}
\label{lncgqqbar}
\end{figure}

As done for the previous channel, let us consider the graphical
representation of this channel in the large-$N_c$ limit in
Fig.~\ref{lncgqqbar}. After replacing the gluon line with a
quark-antiquark pair we are left with two independent fermion
lines which scatter separately with the nucleus. Each of them
resembles the photon emission case and therefore we expect, even
before performing the calculation, to obtain a convolution of two
Fourier transforms of the dipole cross section. In the correlation
limit, the two terms in (\ref{ggtoqqbar}) have a simple
explanation in terms of a hard scattering. The first term accounts
for the cases where the hard scattering involves only one of the
two quark lines, while the second term is an interference term
when the large momentum transfer involves the two participants.

This channel had already been considered in \cite{Blaizot:2004wv}
where, due to the choice of gauge, the separation of the amplitude
in terms of splitting function and multiple scattering terms is
not visible. It is possible to show our expressions above are
consistent with their results when expressed in the same set of
coordinates and momentum variables.

\subsubsection{g$\to$gg}

In order to study the partonic process
$g\to gg$, we need to derive the splitting function first. It can
be written in momentum space as
\begin{equation}
\Psi_{g\to gg}(z, p_{\perp})=\frac{1}{\sqrt{8p^+ k_1^+ k_2^+}}\frac{V_{g\to gg}}{k_1^- +k_2^- - p^-},
\end{equation}
where $V_{g\to gg}$ is just the three-gluon vertex with the coupling and color factor factorized out. This can be written as
\begin{equation}
V_{g\to gg} = \epsilon_a^{\alpha}\epsilon_b^{\beta}\epsilon_c^{\gamma}\left[g_{\alpha \beta} \left(p_a-p_b\right)_{\gamma}+g_{ \beta \gamma} \left(p_b-p_c\right)_{\alpha}+g_{\gamma \alpha} \left(p_c-p_a\right)_{\beta}\right].
\end{equation}
Here $\epsilon_i^{\mu}$ represents the polarization vector for gluon $i$ with four momentum $p_i$. It is straightforward to find that
\begin{equation}
\sum_{\textrm{spin}}\left|V_{g\to gg}\right|^2=\frac{8p^2_{\perp}}{z(1-z)}\left[\frac{z}{1-z}+\frac{1-z}{z}+z(1-z)\right].
\end{equation}

After summing over all polarizations, the squared splitting function in transverse coordinate space reads
\begin{equation}
\sum \Psi_{g\to gg}^*(z,u^{\prime})\Psi_{g\to gg}(z,u)=(2\pi)^2 \frac{4}{p^+}\left[\frac{z}{1-z}+\frac{1-z}{z}+z(1-z)\right]\frac{u^{\prime}\cdot u}{u^{\prime 2} u^{ 2}}.
\end{equation}

Now let us turn our attention to the multiple scattering terms. Since all the particles involved in the process are gluons, all terms contain only Wilson lines in the adjoint representation. In the following we give the explicit forms of the scattering terms with their respective large-$N_c$ expressions in terms of fundamental Wilson lines.
\begin{align}
&\left\langle f_{ade}\left[W(x_1)W^\dagger(x'_1)\right]^{db}\left[W(x_2)W^\dagger(x'_2)\right]^{ec}f_{abc}\right\rangle_{x_g}
\simeq \left\langle\text{Tr}U^\dagger(x_1)U(x'_1)\right\rangle_{x_g}\left\langle\text{Tr}U(x_2)U^\dagger(x'_2)\right\rangle_{x_g}\notag\\
&\hspace{9cm}\times\left\langle\text{Tr}U(x_1)U^\dagger(x'_1)U(x'_2)U^\dagger(x_2)\right\rangle_{x_g},
\end{align}
\begin{eqnarray}
\left\langle f_{ade}W^{db}(x_1)W^{ec}(x_2)f_{fbc}W^{fa}(v')\right\rangle_{x_g}&\simeq &
\left\langle\text{Tr}U^\dagger(x_1)U(v')\right\rangle_{x_g}
\left\langle\text{Tr}U(x_2)U^\dagger(v')\right\rangle_{x_g}\nonumber\\
&&~~~\times\left\langle\text{Tr}U(x_1)U^\dagger(x_2)\right\rangle_{x_g}\ ,\\
\left\langle f_{ade}W^{db}(x'_1)W^{ec}(x'_2)f_{fbc}W^{fa}(v)\right\rangle_{x_g}&\simeq &
\left\langle\text{Tr}U^\dagger(v)U(x'_1)\right\rangle_{x_g}
\left\langle\text{Tr}U(v)U^\dagger(x'_2)\right\rangle_{x_g}\nonumber\\
&&~~~\times\left\langle\text{Tr}U(x'_2)U^\dagger(x'_1)\right\rangle_{x_g},\\
N_c\left\langle\text{Tr}W(v)W^\dagger(v')\right\rangle_{x_g}&\simeq& N_c\left\langle\text{Tr}U^\dagger(v)U(v')\right\rangle_{x_g}\left\langle\text{Tr}U(v)U^\dagger(v')\right\rangle_{x_g}.
\end{eqnarray}

The correlation limit is applied by following the procedure developed in the DIS case and reproduced in the $q\to qg$ channel. By inspection of the multiple scattering terms above, it is easy to see that the same kind of cancelations occur for this channel. Since the lowest order terms left after the various cancelations come from the first of the scattering terms, it is easy to see that the final result will involve combinations of one WW distribution and two Fourier transforms of the dipole cross section. After some algebra we arrive at
\begin{eqnarray}
\frac{d\sigma ^{pA\rightarrow ggX}}{d{\cal P.S.}} &=&x_{p}g(x_{p})64\pi ^{3}\frac{\alpha
_{S}^{2}}{P_{\perp }^{4}}z(1-z)\left[
\frac{1-z}{z}+\frac{z}{1-z}+z(1-z)\right] \int \frac{
d^{3}v}{(2\pi )^{3}}\frac{d^{3}v^{\prime }}{(2\pi
)^{3}}e^{-iq_{\perp }\cdot
(v-v^{\prime })}  \nonumber \\
&&\times \left[ \left(
z^{2}+(1-z)^{2}\right)\frac{1}{N_{c}}\left\langle
\text{Tr}U(v)U^{\dagger }(v^{\prime })\right\rangle _{x_{g}}
\left\langle \text{Tr}\left[
F^{i-}(\vec{v})\mathcal{U}^{[+]\dagger }F^{i-}(\vec{v}^{\prime
})\mathcal{U}
^{[-]}\right] \right\rangle _{x_{g}}\right.   \nonumber \\
&& \left.+ \frac{1}{N_{c}}\!\left\langle
\text{Tr}U(v)U^{\dagger }(v^{\prime })\right\rangle _{x_{g}}
\frac{1}{N_{c}}\left\langle \text{Tr}U(v^{\prime})U^{\dagger
}(v)\right\rangle _{x_{g}}\left\langle \text{Tr}\left[
F^{i-}(\vec{v})\mathcal{U}^{[+]\dagger }F^{i-}(\vec{v}^{\prime })\mathcal{U}%
^{[+]}\right] \right\rangle _{x_{g}} \right. \nonumber \\
&& \left. +2z(1-z)\frac{1}{N_{c}}\left\langle
\text{Tr}F^{i-}(\vec{ v})U(v)U^{\dagger }(v^{\prime
})\right\rangle _{x_{g}}\frac{1}{N_{c}}\left\langle \text{Tr}
F^{i-}(\vec{v}^{\prime })U(v^{\prime })U^{\dagger }(v)
\right\rangle _{x_{g}}\right] ,
\end{eqnarray}
which is straightforward to compare to the factorized expression in (\ref{factggg}).

\begin{figure}[tbp]
\begin{center}
\includegraphics[width=9cm]{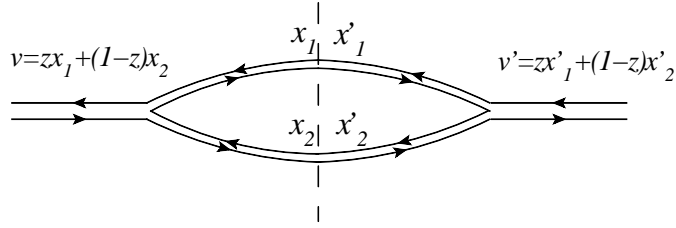} \\[0pt]
\end{center}
\caption[*]{Graphical representation of the splitting $g\to gg$ in the large-$N_c$ limit, in the amplitude and the conjugate amplitude.}
\label{lncggg}
\end{figure}

This structure could have been anticipated by looking at the graphical representation of this process in the large-$N_c$ limit shown in Fig. \ref{lncggg}. In terms of the hard scattering picture used in previous sections the structure of the expression above is consistent with previous results. The first and third term look exactly the same as the two terms in the $g\to q\bar{q}$ case and they correspond to the case in which the hard scattering does not involve the inner loop in Fig. \ref{lncggg}. The second term corresponds to the case where the hard scattering occurs in the inner loop. It has the same structure as one of the terms found in the $q\to qg$ case with an additional convolution associated to the extra quark line in the top of the diagram.

\section{Conclusion}

In this paper, we have studied and established an effective $k_t$-factorization for dijet production at small-$x$ in dilute-dense collisions. Although $k_t$-dependent parton distributions are different in different processes, they can be calculated and related to each other. We found that there are two fundamental unintegrated gluon distributions, namely, $xG^{(1)}$ and $xG^{(2)}$, at small $x$. Although other different gluon distributions appear in many different dijet production processes, one can compute them and find that they are related to these two fundamental gluon distributions in the large $N_c$ limit. In terms of the CGC framework, this means that the two- and four-point functions are enough to compute any dijet cross section, in the small momentum imbalance limit. In addition, there is similar conclusion for the quark distributions at small $x$
\cite{Xiao:2010sp}. By doing so, we can restore the predictive power of the theory.

Therefore, as part of the conclusion, we would like to summarize the empirical rules in the large-$N_c$ limit\footnote{Note that one does not need to take the large $N_c$ limit in the calculation of $xG^{(1)}$ and $xG^{(2)}$ in DIS dijet and photon-jet in pA collisions, respectively. However, the large $N_c$ limit is essential in order to eliminate other non-universal distributions or correlators in other different dijet channels, i.e., $qg\to qg$, $gg\to q\bar q$ and $gg\to gg$ in pA collisions. } in this effective $k_t$-factorization for dilute-dense system as follows:
\begin{itemize}
\item The cross section can be still separated into the products of the hard parts and parton distributions;
\item The hard factors should be calculated separately for each individual graph since the parton distribution associated with each graph may be different;
\item By replacing gluons into double lines, transform the Feynman graphs of the hard part into large $N_c$ planar graphs. The planar graphs show that there are only two building blocks, namely, quark lines and color singlet quark loops, in any graphs (see e.g., Figs.~\ref{lncqqg}, \ref{lncgqqbar} and \ref{lncggg});
\item Quark lines always have interactions with the dense target which contribute $xG^{(2)}$ or $F_{x_g}(q_{\perp})$ to the gluon distribution. However, the color singlet quark loop may or may not interact with the dense target due to its peculiar color structure. If the quark loop participates the interaction, it contributes $xG^{(1)}$ to the gluon distribution. \item If there are multiple objects involved in the soft interaction, the resulting gluon distribution can be written in terms of convolutions of all contributions in momentum space. For quark distributions in the dense target, there are similar rules which can be found in Ref.~\cite{Xiao:2010sp}.
\end{itemize}
Using the above rules and calculating the coefficients of gluon
distributions as illustrated in Ref.~\cite{Bomhof:2006dp}, it is
then straightforward to write down cross sections in terms of
products of hard parts and parton distributions as illustrated in
the context of this paper.

There have been ambiguities regarding the unintegrated gluon
distributions for more than a decade. In this paper, we resolve
this decade-long puzzle through explicit operator definitions and
propose measurements in physical processes which probe the
distributions directly. In particular, we find that
quark-antiquark correlation in DIS collisions can probe the
Weizs\"acker-Williams gluon distribution formulated in CGC many
years ago.

It is well-known that in the color dipole approach, the cross
sections of inclusive DIS and SIDIS \cite{Marquet:2009ca} at
small-$x$ are proportional to the dipole cross section. Since the
dipole cross section can be written as Fourier transform of the
normalized gluon distribution $F_x(q^2_{\perp})$, we can study the
unintegrated gluon distribution $xG^{(2)}$ of nuclei through
inclusive DIS and SIDIS at EIC. Moreover, using DIS dijet
(dihadron) processes at EIC in the correlation limit, we can
directly probe the Weizs\"acker-Williams gluon distribution
$xG^{(1)}$ which is the distribution that actually counts the
number of gluons in the nuclear wave function. This would give us
the golden opportunity to access the saturated WW gluon
distribution which has been studied for many years.

Recently, both STAR and PHENIX Collaborations have published
experimental results on di-hadron correlations in $dAu$
collisions, where a strong back-to-back de-correlation of the two
hadrons was found in the forward rapidity region of the
deuteron~\cite{rhic-data}. These results have stimulated a number
of theoretical calculations in the CGC formalism, where different
assumptions have been made in the
formulations~\cite{{Albacete:2010pg},Tuchin:2009nf}, though not
the correlation limit we had to use in the present study. In
particular, the numerical evaluation in
Ref.~\cite{Marquet:2007vb,{Albacete:2010pg}} only contains the
first term in the $qg$ channel in Eq.(B22). The second term, as
well as other missing terms due to the use of a Gaussian
approximation \cite{Dumitru:2010ak}, are equally important and
should be taken into account to interpret the STAR data. In
addition, we also present the first CGC calculations on the $g\to
q\bar q$ and $g\to gg$ channels in $pA$ collisions. Although these
channels are subdominant in the forward dijet productions, they
are important in the central rapidity region.

At RHIC and LHC, by measuring the direct photon-jet correlation in
$pA$ collisions, one can gain direct information about the dipole
unintegrated gluon distribution $xG^{(2)}$. Furthermore, by
investigating the dijet (quark-gluon or gluon-gluon jet)
production in the correlation limit, one can test the universality
of gluon distributions, and begin to see the convolutions of these
two unintegrated gluon distributions. Using dijet production with
more general kinematics, one can even probe deeper the small-$x$
QCD dynamics, as multi-gluon distributions become crucial.

\begin{acknowledgments}
We thank Al Mueller for stimulating discussions and critical
reading of the manuscript. We thank Alberto Accardi, Emil Avsar, Markus Diehl,
Volker Koch, Larry McLerran, Stephane Munier, Jianwei Qiu, Anna Stasto, Raju Venugopalan and Xin-Nian Wang for
helpful conversations. This work was supported in part by the U.S.
Department of Energy under the contracts DE-AC02-05CH11231 and DOE
OJI grant No. DE - SC0002145. We are grateful to RIKEN, Brookhaven
National Laboratory and the U.S. Department of Energy (contract
number DE-AC02-98CH10886) for providing the facilities essential
for the completion of this work. We also thank the Institute for
Nuclear Theory at the University of Washington for its hospitality
and the Department of Energy for partial support during the
completion of this work.
\end{acknowledgments}

\appendix

\section{Derivation of two gluon distributions}

\subsection{The Weizs\"{a}cker-Williams gluon distribution}
\label{wwg}
Let us focus on the Weizs\"{a}cker-Williams gluon distribution $%
xG^{(1)}\left( x,q_{\perp }\right) $ first. Here we provide a derivation of
this gluon distribution from its operator definition, together with the gauge
links for a large nucleus, by using the McLerran-Venugopalan model. According to
its definition
\begin{equation}
xG^{(1)}\left( x,q_{\perp }\right) =2\int \frac{\text{d}\xi _{-}\text{d}%
^{2}\xi _{\perp }e^{-iq_{\perp }\cdot \xi _{\perp }-ixP^{+}\xi _{-}}}{\left(
2\pi \right) ^{3}P^{+}}\left\langle \text{Tr}\left[ F\left( \xi \right)
\mathcal{U}^{\left[ +\right] \dagger }F\left( 0\right) \mathcal{U}^{\left[ +%
\right] }\right] \right\rangle ,
\end{equation}%
with $\mathcal{U}^{[+]}=U^{n}[0,+\infty ;0]U^{n}[+\infty ,\xi _{-};\xi
_{\perp }]$. The definition above is gauge invariant. However, in order to
simplify the calculation, we have chosen to use covariant gauge. Thus, it is then easy to write it as
\begin{eqnarray}
xG^{(1)}(x,q_\perp) &=&2\int \frac{\text{d}\xi _{-}\text{d}%
^{2}\xi _{\perp }e^{-iq_{\perp }\cdot \xi _{\perp }-ixP^{+}\xi _{-}}}{\left(
2\pi \right) ^{3}P^{+}}  \left\langle \text{Tr}\left[ U^{n}[+\infty ,\xi _{-};\xi _{\perp
}]F\left( \xi \right) U^{n\dagger }[+\infty ,\xi _{-};\xi _{\perp
}]\right.\right.\notag \\
&&\times \left.\left. U^{n\dagger }[0,+\infty ;0]F\left( 0\right) U^{n}[0,+\infty ;0]\right]
\right\rangle  \notag \\
&=&\int \frac{\text{d}\xi _{-}\text{d}^{2}\xi _{\perp }e^{-iq_{\perp }\cdot
\xi _{\perp }-ixP^{+}\xi _{-}}}{\left( 2\pi \right) ^{3}P^{+}}  \notag \\
&&\times \left\langle \left[ W^{n\dagger }[+\infty ,\xi _{-};\xi _{\perp
}]_{ab}F^{a}\left( \xi \right) W^{n\dagger }[0,+\infty ;0]_{cb}F^{c}\left(
0\right) \right] \right\rangle ,
\end{eqnarray}%
where $W^{n\dagger }[+\infty ,\xi _{-};\xi _{\perp }]_{ab}$ now is in the
adjoint representation. Following Belitsky \textit{et al} \cite%
{Belitsky:2002sm}, we can insert a complete set of one particle intermediate
states. Notice that for the quark distribution at small-$x$, we need to have
two particle intermediate states due to the antiquark spectator. Therefore, the
gluon distribution reads
\begin{equation}
xG^{(1)}(x,q_{\perp })=\int \frac{\text{d}\xi _{-}\text{d}^{2}\xi _{\perp
}e^{-iq_{\perp }\cdot \xi _{\perp }-ixP^{+}\xi _{-}}}{\left( 2\pi \right)
^{3}P^{+}}\int \frac{d^{4}P^{\prime }}{(2\pi )^{4}}(2\pi )\delta (P^{\prime
2}-m^{2})\mathcal{A}_{a}^{\dagger }(0)\mathcal{A}_{a}(\xi ),
\end{equation}%
where we introduced the amplitude
\begin{equation}
\mathcal{A}_{b}(\xi )\equiv \left\langle P^{\prime }|W^{n\dagger }[+\infty
,\xi _{-};\xi _{\perp }]_{ab}F^{a}\left( \xi \right) |P\right\rangle .
\end{equation}%
This amplitude then takes the form
\begin{equation}
\mathcal{A}_{a}(\xi )=\int \frac{d^{4}k}{(2\pi )^{4}}e^{i\xi _{-}k^{+}+i\xi
_{\perp }k_{\perp }}(2\pi )^{4}\delta ^{(4)}\left( k+P^{\prime }-P\right)
\mathcal{A}_{a}(k).
\end{equation}%
By plugging in everything into the definition, we find
\begin{equation}
xG^{(1)}(x,q_{\perp })=\frac{2}{\left( 2\pi \right) ^{3}\left( 2P^{+}\right)
^{2}}\mathcal{A}_{a}^{\dagger }(x,q_{\perp })\mathcal{A}_{a}(x,q_{\perp }).
\end{equation}
Let us define $\widetilde{\mathcal{A}}_{a}(x^{-},R_{\perp })$ as the Fourier
transform of $\mathcal{A}_{a}(x,q_{\perp })$. Like what we have done for the
quark distributions~\cite{Xiao:2010sp}, we can compute the diagrammatic contributions to $%
\widetilde{\mathcal{A}}_{a}(x_{-},R_{\perp })$ order by order and resum it
in coordinate space which gives
\begin{equation}
\widetilde{\mathcal{A}}_{a}(x_{-},R_{\perp })\!=\!W_{ba}\left( x_{-},R_{\perp
}\right) F_{b}^{+i}\left( R_{\perp }\right)\! =\!-\!\int {d}^{2}x_{\perp }\rho
_{b}\left( x_{-},x_{\perp }\right) W_{ba}\left( x_{-},R_{\perp }\right)
\nabla _{R_{\perp }}G\left( R_{\perp }-x_{\perp }\right) ,
\end{equation}
where%
\begin{equation}
W\left( x_{-},R_{\perp }\right) =\text{\textit{T}}\mathit{\exp }\left[
-ig\int_{x_{-}}^{+\infty }{d}z_{-}\int {d}^{2}z_{\perp }G\left( x_{\perp
}-z_{\perp }\right) \rho _{c}\left( z_{-},z_{\perp }\right) t^{c}\right] ,
\end{equation}%
with $t^{c}$ being the adjoint color matrix. Eventually, one should be able
to write
\begin{align}
&xG^{(1)}(x,q_{\perp })=\frac{1}{4\pi ^{3}}\int {d}z_{-}\int d^{2}R_{\perp
}\int {d}z_{-}^{\prime }\int d^{2}R_{\perp }^{\prime }e^{iq_{\perp }\cdot
\left( R_{\perp }-R_{\perp }^{\prime }\right) +ixP^{+}\left(
x_{-}-x_{-}^{\prime }\right) }\nonumber\\
&\hspace{9cm}\times \widetilde{\mathcal{A}}_{a}^{\dagger
}(z_{-},R_{\perp })\widetilde{\mathcal{A}}_{a}(z_{-}^{\prime },R_{\perp
}^{\prime }).
\end{align}%
In arriving to the expression above, we have put in a normalization factor of $%
2P^{+}$ which comes from $\left\langle P\left\vert \cdots \right\vert
P\right\rangle $. In the following, we need to average the above expression
with a gaussian distribution of color charges as proposed in CGC. Therefore
\begin{eqnarray}
xG^{(1)}(x,q_{\perp }) &=&\frac{1}{4\pi ^{3}}\int d^{2}R_{\perp }\int
d^{2}R_{\perp }^{\prime }e^{iq_{\perp }\cdot \left( R_{\perp }-R_{\perp
}^{\prime }\right) +ixP^{+}\left( z_{-}-z_{-}^{\prime }\right)
}\int_{-\infty }^{+\infty }{d}z_{-}{d}^{2}z_{\perp }\int_{-\infty }^{+\infty
}{d}z_{-}^{\prime }{d}^{2}z_{\perp }^{\prime }  \notag \\
&&\times \left\langle \rho _{b}\left( z^{-},z_{\perp }\right) W_{ba}\left(
z^{-},R_{\perp }\right) \rho _{c}\left( z^{\prime -},z_{\perp }^{\prime
}\right) W_{ca}^{\dagger }\left( z^{\prime -},R_{\perp }^{\prime }\right)
\right\rangle _{\rho }  \notag \\
&&\times \nabla _{R_{\perp }}G\left( R_{\perp }-z_{\perp }\right) \nabla
_{R_{\perp }^{\prime }}G\left( R_{\perp }^{\prime }-z_{\perp }^{\prime
}\right) .
\end{eqnarray}%
Assuming a Gaussian distribution of sources, it is easy to prove that
\begin{eqnarray}
&&\left\langle \rho _{b}\left( z_{-},z_{\perp }\right) W_{ba}\left(
z_{-},R_{\perp }\right) \rho _{c}\left( z_{-}^{\prime },z_{\perp }^{\prime
}\right) W_{ca}^{\dagger }\left( z_{-}^{\prime },R_{\perp }^{\prime }\right)
\right\rangle _{\rho } \nonumber \\
&=&\left\langle \rho _{b}\left( z_{-},z_{\perp }\right) \rho _{c}\left(
z_{-}^{\prime },z_{\perp }^{\prime }\right) \right\rangle \left\langle
W_{ba}\left( z_{-},R_{\perp }\right) W_{ca}^{\dagger }\left( z_{-}^{\prime
},R_{\perp }^{\prime }\right) \right\rangle _{\rho } \\
&=&\delta _{bc}\delta \left( z_{-}-z_{-}^{\prime }\right) \mu ^{2}\left(
z_{-},z_\perp-z'_\perp\right) \left\langle W_{ba}\left( z_{-},R_{\perp }\right)
W_{ca}^{\dagger }\left( z_{-}^{\prime },R_{\perp }^{\prime }\right)
\right\rangle _{\rho },
\end{eqnarray}
by using the fact that $t_{ab}^{c}=-if_{abc}$(Note that $f_{abc}$ is
anti-symmetric. See~\cite{Iancu:2002xk} for more details). Therefore,
we have
\begin{eqnarray}
xG^{(1)}(x,q_{\perp }) &=&\frac{g^{2}}{4\pi ^{3}}\int d^{2}R_{\perp }\int
d^{2}R_{\perp }^{\prime }e^{iq_{\perp }\cdot \left( R_{\perp }-R_{\perp
}^{\prime }\right) }\int_{-\infty }^{+\infty }{dz_{-}d}^{2}z_{\perp }d^2z'_\perp\mu
^{2}\left( z_{-},z_\perp-z'_\perp\right) \text{ }  \notag \\
&&\times \text{Tr}\left\langle W\left( z_{-},R_{\perp }\right) W^{\dagger
}\left( z_{-},R_{\perp }^{\prime }\right) \right\rangle \nabla _{R_{\perp
}}G\left( R_{\perp }-z_{\perp }\right) \nabla _{R_{\perp }^{\prime }}G\left(
R_{\perp }^{\prime }-z'_{\perp }\right).\label{G1GG}
\end{eqnarray}
The different factors in the integral above can all be written in terms of a single function when a Gaussian distribution of sources is used. Let
\begin{align}
\Gamma(R_\perp-R'_\perp)=&\;g^4\int d^2z_\perp d^2z^{\prime}_\perp dz_-\mu^2(z_-,z_\perp-z'_\perp)\notag\\
&\times\left[G(R_\perp-z_\perp)-G(R'_\perp-z_\perp)\right]\left[G(z^{\prime}_\perp-R_\perp)-G(z^{\prime}_\perp-R'_\perp)\right].
\end{align}
In terms of $\Gamma$, Eq.~(\ref{G1GG}) takes the form
\begin{equation}
xG^{(1)}(x,q_\perp)=\frac{S_{\perp }}{16\pi ^{4}\alpha _{s}}\frac{N_{c}^{2}-1}{N_{c}}\int d^{2}r_{\perp }e^{iq_{\perp }\cdot r_{\perp }}\frac{\nabla^2\Gamma(r_\perp)}{\Gamma(r_\perp)}\left[1-\exp \left( -\frac{N_c}{2}\Gamma(r_\perp)\right) \right] .
\end{equation}
In particular, for the McLerran-Venugopalan model, the function $\Gamma$ can be evaluated explicitly giving the well-known result
\begin{equation}
xG^{(1)}(x,q_\perp)=\frac{S_{\perp }}{4\pi ^{4}\alpha _{s}}\frac{N_{c}^{2}-1}{N_{c}}\int d^{2}r_{\perp }e^{iq_{\perp }\cdot r_{\perp }}\frac{1}{r_{\perp }^{2}}\left[1-\exp \left( -\frac{1}{4}r_{\perp }^{2}Q_{s}^{2}\right) \right] ,
\end{equation}
where $Q_{s}^{2}$ is the gluon saturation scale.

\subsection{The dipole gluon distribution}
\label{dipoleg}
According to its definition
\begin{equation}
xG^{\left( 2\right) }(x,q_{\perp })=2\int \frac{\text{d}\xi _{-}\text{d}%
^{2}\xi _{\perp }e^{-iq_{\perp }\cdot \xi _{\perp }-ixP^{+}\xi _{-}}}{\left(
2\pi \right) ^{3}P^{+}}\left\langle \text{Tr}\left[ F\left( \xi \right)
\mathcal{U}^{\left[ -\right] \dagger }F\left( 0\right) \mathcal{U}^{\left[ +%
\right] }\right] \right\rangle ,
\end{equation}%
with $\mathcal{U}^{[+]}=U^{n}[0,+\infty ;0]U^{n}[+\infty ,\xi _{-};\xi
_{\perp }]$ and $\mathcal{U}^{[-]}=U^{n}[0,-\infty ;0]U^{n}[-\infty ,\xi
_{-};\xi _{\perp }]$. Here we have chosen the covariant gauge to do the
calculation.\ Thus, one gets
\begin{eqnarray}
xG^{\left( 2\right) }(x,q_{\perp }) &=&2\int \frac{\text{d}\xi _{-}\text{d}%
^{2}\xi _{\perp }e^{-iq_{\perp }\cdot \xi _{\perp }-ixP^{+}\xi _{-}}}{\left(
2\pi \right) ^{3}P^{+}}  \notag \\
&&\times \left\langle \text{Tr}\left[
%\begin{array}{c}
U^{n}[+\infty ,\xi _{-};\xi _{\perp }]F\left( \xi \right) U^{n\dagger
}[-\infty ,\xi _{-};\xi _{\perp }] \right.\right.\notag\\
%\\
%\times
&&\left.\left.\qquad \quad U^{n\dagger }[0,-\infty ;0]F\left( 0\right) U^{n}[0,+\infty ;0]%
%\end{array}%
\right] \right\rangle .
\end{eqnarray}%
By inserting the intermediate state and replacing the $\left\langle
P\left\vert \cdots \right\vert P\right\rangle $ by the ensemble average $%
\left\langle \cdots \right\rangle _{\rho }$, we get
\begin{equation}
xG^{\left( 2\right) }(x,q_{\perp })=\frac{1}{2\pi ^{3}}\left\langle \text{Tr}%
\left[ \mathcal{B}^{\dagger }(x,q_{\perp })\mathcal{B}(x,q_{\perp })\right]
\right\rangle _{\rho },
\end{equation}%
with
\begin{equation}
\mathcal{B}(\xi )\equiv \left\langle P^{\prime }| U^{n}[+\infty ,\xi _{-};\xi _{\perp }]F\left( \xi
\right) U^{n\dagger }[-\infty ,\xi _{-};\xi _{\perp }]|P\right\rangle.
\end{equation}%
and
\begin{equation}
\mathcal{B}(\xi )=\int \frac{d^{4}k}{(2\pi )^{4}}e^{i\xi _{-}k^{+}+i\xi
_{\perp }k_{\perp }}(2\pi )^{4}\delta ^{(4)}\left( k+P^{\prime }-P\right)
\mathcal{B}(k)
\end{equation}%
In CGC, we can find that, in covariant gauge, the only non-trivial field strength
is $F_{a}^{+i}\left( x_{\perp }\right) =-\partial ^{i}\int $d$^{2}y_{\perp
}G\left( x_{\perp }-y_{\perp }\right) \rho _{a}\left( x_{-},y_{\perp
}\right) $. Therefore, if we write
\begin{eqnarray}
\mathcal{B}(q) &=&\int \text{d}x_{-}\text{d}^{2}R_{\perp }e^{iR_{\perp
}q_{\perp }}\widetilde{\mathcal{B}}(x_{-},R_{\perp }) \\
&=&\int \text{d}x_{-}\text{d}^{2}R_{\perp }e^{iR_{\perp }q_{\perp
}}U^{n}[+\infty ,x_{-};R_{\perp }]F\left( R_{\perp }\right) U^{n\dagger
}[-\infty ,x_{-};R_{\perp }],
\end{eqnarray}%
we can easily see that $\mathcal{B}(q)\propto \partial ^{i}U^{n\dagger
}[-\infty ,+\infty ;R_{\perp }].$ In the above derivation, we have to
assume that $e^{ix^{-}k^{+}}\simeq 1$. This can be justified by noting that $%
x^{-}$ is integrated from $-L$ to $+L$ with $L$ being the longitudinal width
of the nucleus. For small $k^{+}=xP^+$ with small $x$, we have $k^{+}L\ll 1$.

Eventually, one gets%
\begin{eqnarray}
xG^{\left( 2\right) }(x,q_{\perp }) &=&\frac{N_{c}}{2\pi ^{3}}\int
d^{2}R_{\perp }\int d^{2}R_{\perp }^{\prime }e^{iq_{\perp }\cdot \left(
R_{\perp }-R_{\perp }^{\prime }\right) }\frac{\nabla _{R_{\perp }}\cdot
\nabla _{R_{\perp }^{\prime }}}{g^{2}}\frac{1}{N_{c}}\text{Tr}\left[ \langle
U\left( R_{\perp }\right) U^{\dagger }\left( R_{\perp }^{\prime }\right)
\rangle _{\rho }\right]   \notag \\
&=&\frac{q_{\perp }^{2}N_{c}}{2\pi ^{2}\alpha _{s}}S_{\perp
}F_{x_{g}}^{g}(q_{\perp }^{2}) \\
\text{ with }F_{x_{g}}^{g}(q_{\perp }^{2}) &=&\int \frac{d^{2}r_{\perp }}{%
\left( 2\pi \right) ^{2}}e^{iq_{\perp }\cdot r_{\perp }}\frac{1}{N_{c}}\text{%
Tr}\langle U\left( r_{\perp }\right) U^{\dagger }\left( 0\right) \rangle
_{\rho }\simeq \frac{1}{\pi Q_{sq}^{2}}\exp \left[ -\frac{q_{\perp }^{2}}{%
Q_{sq}^{2}}\right] ,
\end{eqnarray}%
where $Q_{sq}^{2}=\frac{\mu _{s}^{2}}{2\pi }\ln \frac{1}{r_{\perp
}^{2}\lambda ^{2}}$ and $\mu _{s}^{2}=\frac{g^{2}}{2}C_{F}\int $d$x^{-}\mu
^{2}\left( x^{-}\right) $. Here the saturation scale $Q_{sq}^{2}$ is
obtained from the fundamental representation and it is usually called quark
saturation momentum.

\section{Evaluations of Correlators}
Here in this section, we summarize the evaluation of the correlators in CGC used above in the main context of the paper.
\subsection{The evaluation of two point functions and $\langle\textrm{Tr}\mathcal{U}^{[\Box]}\rangle$}
First of all, as derived in Ref.~\cite{Gelis:2001da}, for an arbitrary single Wilson line (start at $a^-$ and end at $b^-$) , one can get
\begin{equation}
\langle U\left(a^-,b^-|x_{\perp}\right)\rangle=\exp\left[-\frac{g^4C_F}{2}\int_{a^-}^{b^-}dz^- \mu^{2}(z^-)\int d^2 z_{\perp} G^2(x_{\perp}-z_{\perp})\right].
\end{equation}
Using this result, it is easy to derive that
\begin{equation}
\langle U\left(a^-,b^-|x_{\perp}\right) U\left(b^-,c^-|x_{\perp}\right)\rangle=\langle U\left(a^-,c^-|x_{\perp}\right)\rangle. \label{add}
\end{equation}
The derivation is based on time ordering of $z^-$ and pairing of two adjacent operators.

Furthermore, for two infinite Wilson lines of different transverse position, one gets
\begin{eqnarray}
\langle U\left(x_{\perp}\right)U^{\dagger}\left(y_{\perp}\right)\rangle&=&\exp\left[-\frac{g^4C_F}{2}\int_{-\infty}^{+\infty}dz^- \mu^{2}(z^-)\int d^2 z_{\perp} \left(G(x_{\perp}-z_{\perp})-G(y_{\perp}-z_{\perp})\right)^2\right]  , \nonumber \\
&\simeq&\exp\left[-\frac{g^4C_F}{16\pi}\left(x_{\perp} - y_{\perp}\right)^2\ln\frac{1}{\lambda^2\left(x_{\perp} - y_{\perp}\right)^2}\int_{-\infty}^{+\infty}dz^- \mu^{2}(z^-) \right],
\end{eqnarray}
where $1/\lambda$ stands for the cut-off in the integral. Thus, usually one writes $\langle U\left(x_{\perp}\right)U^{\dagger}\left(y_{\perp}\right)\rangle \simeq \exp\left[-\frac{Q_s^2(x_{\perp}-y_{\perp})^2}{4}\right]$ with $Q_s^2=\frac{g^4C_F}{4\pi}\ln\frac{1}{\lambda^2\left(x_{\perp} - y_{\perp}\right)^2}\int_{-\infty}^{+\infty}dz^- \mu^{2}(z^-)$.

Now we are ready to evaluate $\langle\textrm{Tr}\mathcal{U}^{[\Box]}\rangle$ and show that it is the same as the correlator of two infinite Wilson lines. According to the definition, one can easily find that
\begin{eqnarray}
\langle\textrm{Tr}\mathcal{U}^{[\Box]}\rangle&=&\langle\textrm{Tr}\left[U\left(0,+\infty|0_{\perp}\right)U\left(+\infty,\xi^-|\xi_{\perp}\right)U\left(\xi^-,-\infty|\xi_{\perp}\right)U\left(-\infty,0|0_{\perp}\right)\right]\rangle \\
&=& \langle \textrm{Tr}\left[ U\left(0_{\perp}\right)U^{\dagger}\left(\xi_{\perp}\right)\right]\rangle
\end{eqnarray}
where we have used Eq.~(\ref{add}). Therefore, one can easily relate $\frac{1}{N_c}\langle\textrm{Tr}\mathcal{U}^{[\Box]}\rangle$ to $F_x\left(q_{\perp}^2\right)$ through Fourier transform.

\subsection{Evaluation of the 4-point function with a Gaussian distribution of sources}
\label{4point}
The derivation presented here follows closely the method presented in \cite{Blaizot:2004wv} and used also in \cite{Dominguez:2008aa} where other 4-point functions have been calculated. For the sake of completeness and to make the presentation self-contained we will briefly review how to calculate the 4-point function $S^{(4)}_{x_g}(x_1,x_2;x_2^{\prime},x_1^{\prime})=\frac{1}{N_c}\left\langle\text{Tr}U(x_1)U^\dagger(x_1^{\prime})U(x_2^{\prime})U^\dagger(x_2)\right\rangle_{x_g}$ for a Gaussian distribution of charges. Details of the general procedure are given in \cite{Blaizot:2004wv}.

The nuclear average of a function of the gauge field is defined by
\begin{equation}
\langle f[A]\rangle_{x_g}=\int\mathcal{D}\rho\exp\left\{-\int d^2x\,d^2y\,dz^+\,\frac{\rho_c(z^+,x)\rho_c(z^+,y)}{2\mu_{x_g}^2(z^+)}\right\}f[A],
\end{equation}
where the color charge $\rho$ and the gauge field are related by
\begin{equation}
-\nabla_\perp^2A^-_c(z^+,x)=g_S\rho_c(z^+,x)\:,
\end{equation}
and $\mu_{x_g}^2(z^+)$ is the density of color charges at a given $z^+$. This Gaussian distribution allows us to express any correlator in terms of the elementary correlator of two color charges
\begin{equation}
\langle\rho_c(z^+,x)\rho_d(z^{\prime+},y)\rangle_{x_g}=\delta_{cd}\delta(z^+-z^{\prime+})\delta^{(2)}(x-y)\mu_{x_g}^2(z^+).\label{elemcorr}
\end{equation}

In order to do this, the Wilson lines must be expanded in terms of $g_S\rho$ and then apply Wick's theorem. The Wilson lines are naturally expanded in terms of the gauge field and not the color charge density, therefore it is useful to express the elementary correlator (\ref{elemcorr}) in terms of the gauge field.
\begin{equation}
g_S^2\langle A_c^-(z^+,x)A_d^-(z^{\prime+},y)\rangle_{x_g}=\delta_{cd}\delta(z^+-z^{\prime+})\mu_{x_g}^2(z^+)L_{xy},
\end{equation}
with $L$ given in terms of the two-dimensional massless propagator $G_0$,
\begin{equation}
L_{xy}=g_S^4\int d^2z\;G_0(x-z)G_0(y-z),\qquad G_0(x)=\int\frac{d^2k}{(2\pi)^2}\frac{e^{ik\cdot x}}{k^2}.
\end{equation}
\begin{figure}[tbp]
\begin{center}
\includegraphics[width=15cm]{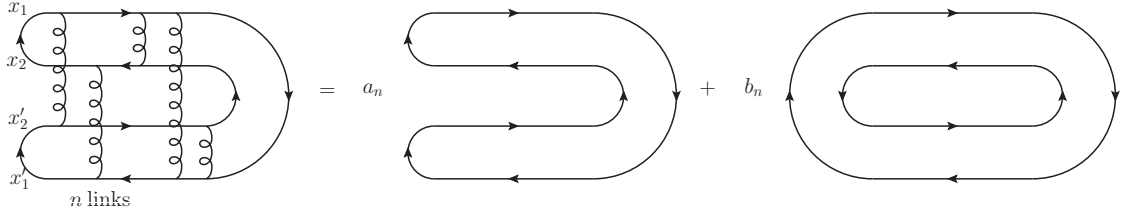}
\end{center}
\caption[*]{Graphical representation of the terms in the series expansion of the 4-point function.}
\label{nlinks}
\end{figure}

This correlation between two fields has the color structure of a gluon link. This, together with the locality of the correlator in the $z^+$ variable, allows for a graphical representation of each of the terms of the expansion of the Wilson lines. For the particular color structure of the 4-point function we are interested in diagrams which look like the left hand side of Fig. \ref{nlinks}. One kind of contribution from these diagrams that is easy to evaluate is the contribution from links with both ends attached to the same line. For a Wilson line at a transverse coordinate $x$ this kind of link gives a factor of $-C_F\mu^2L_{xx}/2$, it has a color singlet structure and therefore factors out in the evaluation of any specific diagram. When multiple insertions of these particular links are taken into account they can be resummed into the factor
\begin{equation}
T=e^{-\frac{C_F}{2}\mu^2(L_{x_1x_1}+L_{x_2x_2}+L_{x'_1x'_1}+L_{x'_2x'_2})},
\end{equation}
where the contributions from the four Wilson lines involved in the correlator have been included. After factoring out the so-called tadpole contributions we are left with diagrams in which all gluon links connect different Wilson lines. The strategy to evaluate this diagrams is to include the proper $L$ factors for each of the gluon links and to resolve the color structure by means of the Fierz identity $T^a_{ij}T^a_{kl}=\frac{1}{2}\delta_{il}\delta_{jk}-\frac{1}{2N_c}\delta_{ij}\delta_{kl}$. The resummation then is not a trivial task since each diagram will end up being written as a linear combination of the two topologies shown in the right hand side of Fig. \ref{nlinks}. As shown in \cite{Blaizot:2004wv} this difficulty can be overcome by grouping diagrams according to the number of gluon links and then using an inductive procedure to find the value of the $n$th term in the series. By explicitly resolving the $n$th link and using the notation in Fig. \ref{nlinks} we have,
\begin{equation}
\left(\begin{matrix}a_n \\ b_n\end{matrix}\right)=\mu_{x_g}^2(z_n^+)M\left(\begin{matrix}a_{n-1} \\ b_{n-1}\end{matrix}\right),
\end{equation}
where the matrix $M$ is given by all the different ways in which the $n$th link can be attached,
\begin{equation}
M=\left(\begin{matrix}C_F(L_{x_1x_2}+L_{x'_2x'_1})+\frac{1}{2N_c}F(x_1,x_2;x'_2,x'_1) & -\frac{1}{2}F(x_1,x'_1;x'_2,x_2)\\ -\frac{1}{2}F(x_1,x_2;x'_2,x'_1) & C_F(L_{x_1x'_1}+L_{x'_2x_2})+\frac{1}{2N_c}F(x_1,x'_1;x'_2,x_2)\end{matrix}\right),
\end{equation}
with $F(x_1,x_2;x'_2,x'_1)=L_{x_1x'_2}-L_{x_1x'_1}+L_{x_2x'_1}-L_{x_2x'_2}$. It is easy to solve this recursion relation taking into account the initial condition $a_0=1,b_0=0$. Formally, the solution reads
\begin{equation}
\left(\begin{matrix}a_n \\ b_n\end{matrix}\right)=\left[\prod_{i=1}^n\mu_{x_g}^2(z^+_i)\right]M^n\left(\begin{matrix}1\\ 0\end{matrix}\right).
\end{equation}
In order to find an explicit solution we have to find the eigenvalues $\lambda_\pm$ and eigenvectors of $M$. The solution then takes the form
\begin{equation}
\left(\begin{matrix}a_n \\ b_n\end{matrix}\right)=\left[\prod_{i=1}^n\mu_{x_g}^2(z^+_i)\right]\left(\begin{matrix}a_+\lambda_+^n+a_-\lambda_-^n \\ b_+\lambda_+^n+b_-\lambda_-^n\end{matrix}\right),
\end{equation}
with
\begin{align}
&\lambda_\pm=\left(\frac{N_c}{4}-\frac{1}{2N_c}\right)(L_{x_1x_2}+L_{x'_2x'_1})+\frac{N_c}{4}(L_{x_1x'_1}+L_{x'_2x_2})+\frac{1}{N_c}F(x_1,x_2;x'_2,x'_1)\pm\frac{N_c}{4}\sqrt{\Delta}\;,\\
&a_\pm=\frac{\sqrt{\Delta}\pm F(x_1,x'_2;x_2,x'_1)}{2\sqrt{\Delta}},\qquad b_\pm=\mp\frac{F(x_1,x_2;x'_2,x'_1)}{N_c\sqrt{\Delta}},\\
&\Delta=F^2(x_1,x'_2;x_2,x'_1)+\frac{4}{N_c^2}F(x_1,x_2;x'_2,x'_1)F(x_1,x'_1;x'_2,x_2).
\end{align}

With this result we can now easily resum the contribution from all the diagrams. The 4-point function is then given by
\begin{equation}
S^{(4)}_{x_g}(x_1,x_2;x'_2,x'_1)=\frac{T}{N_c}\sum_{n=0}^\infty\int_{z_1^+<\cdots<z_n^+}\left[N_ca_n(z_1^+,\dots,z_n^+)+N_c^2b_n(z_1^+,\dots,z_n^+)\right].
\end{equation}
When written in terms of the eigenvalues above this expression can be resummed into
\begin{equation}
S^{(4)}_{x_g}(x_1,x_2;x'_2,x'_1)=\frac{T}{N_c}\left[N_c\left(a_+e^{\mu^2\lambda_+}+a_-e^{\mu^2\lambda_-}\right)+N_c^2\left(b_+e^{\mu^2\lambda_+}+b_-e^{\mu^2\lambda_-}\right)\right].
\end{equation}
Using the explicit values shown above, this expression takes the form
\begin{align}
S_{x_g}^{(4)}(x_1,x_2;x'_2,x'_1)=&\;e^{-\frac{C_F}{2}[\Gamma(x_1-x_2)+\Gamma(x'_2-x'_1)]}\nonumber\\
&\times\left[\left(\frac{\sqrt{\Delta}+F(x_1,x'_2;x_2,x'_1)}{2\sqrt{\Delta}}-\frac{F(x_1,x_2;x'_2,x'_1)}{\sqrt{\Delta}}\right)e^{\frac{N_c}{4}\mu^2\sqrt{\Delta}}\right.\notag\\
&\left.+\left(\frac{\sqrt{\Delta}-F(x_1,x'_2;x_2,x'_1)}{2\sqrt{\Delta}}+\frac{F(x_1,x_2;x'_2,x'_1)}{\sqrt{\Delta}}\right)e^{-\frac{N_c}{4}\mu^2\sqrt{\Delta}}\right]\notag\\
&\times e^{-\frac{N_c}{4}\mu^2F(x_1,x'_2;x_2,x'_1)+\frac{1}{2N_c}\mu^2F(x_1,x_2;x'_2,x'_1)},
\end{align}
with $\Gamma(x-y)=\mu^2(L_{xx}+L_{yy}-2L_{xy})$.

Taking the large-$N_c$ limit of this result we find a much simpler expression,
\begin{align}
S_{x_g}^{(4)}(x_1,x_2;x'_2,x'_1)\!\simeq& e^{-\frac{C_F}{2}[\Gamma(x_1-x_2)+\Gamma(x'_2-x'_1)]}\notag\\
&\!-\!\!\frac{F(x_1,x_2;x'_2,x'_1)}{F(x_1,x'_2;x_2,x'_1)}\left(e^{-\frac{C_F}{2}[\Gamma(x_1-x_2)+\Gamma(x'_2-x'_1)]}-e^{-\frac{C_F}{2}[\Gamma(x_1-x'_1)+\Gamma(x'_2-x_2)]}\right).
\end{align}

Note that even this large-$N_c$ result can not be expressed as a product of 2-point functions. The two terms appearing above have a simple interpretation in terms of multiple scatterings that allows us to label the first term as the elastic part and the second term as the inelastic part (see \cite{JalilianMarian:2004da} where the same scattering factor appears in the context of two-gluon production). Taking into account that the Gaussian distribution of sources is equivalent to the two gluon exchange approximation with independent scattering centers, it is easy to see that the first term is what you would expect if only interactions that don't break up nucleons in the nucleus were allowed. In that scenario the quark-antiquark pair is always on a singlet state and the interaction factor can be written as the product of the interaction term for a dipole in the amplitude times an interaction term for the dipole in the conjugate amplitude. The inelastic part takes into account all the interactions were at least one of the nucleons is broken apart, in which case the quark-antiquark pair goes from a singlet state to an octet state. In the large-$N_c$ limit transitions from the octet state to the singlet state are suppressed and therefore the pair remains in the octet state for the rest of the interaction with the nucleus.

%\begin{references}

\end{document}